\documentclass[preprint2]{aastex}

\shorttitle{Empirical Models for Dark Matter Halos. I.}
\shortauthors{Merritt et al.}

\begin{document}

\title{Empirical Models for Dark Matter Halos. I. 
Nonparametric Construction of Density Profiles and Comparison
with Parametric Models}

\author{David Merritt}
\affil{Department of Physics, Rochester Institute of Technology, 
85 Lomb Memorial Drive, Rochester, NY 14623, USA.}

\author{Alister W.\ Graham}
\affil{Mount Stromlo and Siding Spring Observatories, Australian National 
University, Private Bag, Weston Creek PO, ACT 2611, Australia.}

\author{Ben Moore}
\affil{University of Zurich, Winterthurerstrasse 190, CH-8057, 
Z\"urich, Switzerland.}

\author{J\"urg Diemand}
\affil{Department of Astronomy and Astrophysics, University 
of California, 1156 High Street, Santa Cruz, CA 95064, USA.}

\author{Bal{\v s}a Terzi\'c}
\affil{Department of Physics, Northern Illinois University, 
DeKalb, IL 60115, USA.}

\begin{abstract}

We use techniques from nonparametric function estimation theory to
extract the density profiles, and their derivatives, from a set of
$N$-body dark matter halos.
We consider halos generated from $\Lambda$CDM simulations
of gravitational clustering, as well as isolated, spherical collapses.
The logarithmic density slopes $\gamma\equiv d\log\rho/d\log r$ of the
$\Lambda$CDM halos are found to vary as power-laws in radius,
reaching values of $\gamma\approx -1$ at the innermost resolved
radii, $\sim 10^{-2}r_{vir}$.
This behavior is significantly different from that of broken 
power-law models like the NFW profile, but similar to that of 
models like de Vaucouleurs'.
Accordingly, we compare the $N$-body density profiles
with various parametric models to find which provide the best fit.
We consider an NFW-like model with arbitrary inner slope;
Dehnen \& McLaughlin's anisotropic model; Einasto's model
(identical in functional form to S\'ersic's model but
fit to the space density);
and the density model of Prugniel \& Simien that was designed to match 
the deprojected form of S\'ersic's $R^{1/n}$ law.
Overall, the best-fitting model to the $\Lambda$CDM halos
is Einasto's,
although the Prugniel-Simien and Dehnen-McLaughlin models
also perform well.
With regard to the spherical collapse halos, both the
Prugniel-Simien and Einasto models describe the density
profiles well, with an rms scatter some four times smaller than
that obtained with either the NFW-like model or the 3-parameter
Dehnen-McLaughlin model.  
Finally, we confirm recent claims of a
systematic variation in profile shape with halo mass. 
\end{abstract}


\section{Introduction}

A fundamental question is the distribution of matter in bound
systems (galaxies, galaxy clusters, dark matter halos) that form 
in an expanding universe.
Early work on the self-similar collapse of (spherical) primordial
overdensities resulted in virialized structures having density
profiles described by a single power law (e.g., Fillmore \& Goldreich
1984; Bertschinger 1985; Hoffman 1988).  
Some of the first $N$-body simulations were simple cold
collapse calculations like these
(e.g.\ van Albada 1961; Aarseth 1963; H\'enon 1964; Peebles 1970). 
It was quickly realized that, given
appropriately low but non-zero levels of initial random velocity, the
end state of such systems departed from a simple power law,
resembling instead the de Vaucouleurs (1948) $R^{1/4}$ profiles 
observed in elliptical galaxies (e.g.\ van Albada 1982; Aguilar \& Merritt 
1990).
As $N$-body techniques improved, the logarithmic profile slopes 
of cold dark matter (CDM) halos, simulated in hierarchical merger models, 
were also observed to steepen with increasing radius (e.g., West,
Dekel \& Oemler 1987; Frenk et al.\ 1988; Efstathiou et al.\ 1988).
Dubinski \& Carlberg (1991) adopted Hernquist's (1990) double
power law model (itself a modification of Jaffe's 1983 model)
to describe these density profiles.  
This empirical model has an inner logarithmic slope of $-1$ and an outer
logarithmic slope of $-4$.  It was introduced as an analytical
approximation to the deprojected form of de Vaucouleurs' (1948)
profile.  Navarro, Frenk, \& White (1995) modified this to give
the so-called NFW model that has an outer logarithmic slope of $-3$
rather than $-4$, while Moore et al.\ (1998, 1999) suggested that a
further variation having an inner logarithmic slope of $-1.4$ or
$-1.5$ may be more appropriate.


The density profiles of $N$-body halos typically span only
$\sim 2$ decades in radius, between the virial radius and
an inner limit set by the $N$-body resolution.
It has long been clear that other functional forms might fit
such limited data as well or better than the NFW or Moore profiles.
Recently, Navarro et al.\ (2004) argued for a model, like de 
Vaucouleurs', in which the logarithmic slope varies continuously
with radius:
\begin{equation}
{d\ln\rho\over d\ln r} = -2\left({r\over r_{-2}}\right)^\alpha
\label{eq:N03}
\end{equation}
i.e.
\begin{equation}
\rho(r)\propto \exp\left(-A r^\alpha \right),
\label{eq:SerDen0}
\end{equation}
where $r_{-2}$ is the radius at which the logarithmic slope of
the density is $-2$ and $\alpha$ is a parameter  describing
the degree of curvature of the profile.
Merritt et al. (2005) pointed out that this is the same relation
between slope and radius that defines Sersic's (1963, 1968) law, with
the difference that Sersic's law is traditionally applied to the
projected (surface) densities of galaxies, not to the space density.
Merritt et al. further showed that the (space) density profiles of a 
sample of $N$-body halos were equally well fit by 
equation~(\ref{eq:SerDen0}),
or by a deprojected Sersic profile, and that both of these models
provide better fits than an NFW-like, double power-law model with a 
variable inner slope.
Hence, S\'ersic's law -- the function that is so successful at 
describing the luminosity profiles of early-type galaxies and bulges
(e.g. Caon, Capaccioli, \& D'Onofrio 1993; 
Graham \& Guzm\'an 2003, and references therein),
and the projected density of hot gas in galaxy clusters
(Demarco et al. 2003) -- is also an excellent description of
$N$-body halos.

(To limit confusion, we will henceforth refer to equation~(\ref{eq:SerDen0})
as ``Einasto's $r^{1/n}$ model'' when applied to space density profiles,
and as ``S\'ersic's $R^{1/n}$ model'' when applied to projected density
profiles, with $R$ the radius on the plane of the sky.
The former name acknowledges Einasto's (1965, 1968, 1969)
early and extensive use of equation~(\ref{eq:SerDen0}) to model the 
light and mass 
distributions of galaxies (see also Einasto \& Haud 1989).
In addition, we henceforth replace the exponent $\alpha$
by $1/n$ in keeping with the usage established by
S\'ersic and de Vaucouleurs.)

In this paper, we continue the analysis of alternatives to the
NFW and Moore profiles, using a new set of $N$-body halos.
Among the various models that we consider is the Prugniel-Simien
(1997) law, first developed as an analytic approximation
to the deprojected form of the S\'ersic $R^{1/n}$ profile.
Apart from the work of Lima Neto et al.\ (1999), Pignatelli \&
Galletta (1999), and M\'arquez et al.\ (2000, 2001), the
Prugniel-Simien model has received little attention to date.  
Demarco et al.\ (2003) have however applied it to the gas density profiles of
24 galaxy clusters observed with ROSAT, and Terzi\'c \& Graham (2005)
showed that it provides a superior description of the density profiles
of real elliptical galaxies compared with either the Jaffe or Hernquist 
models. 
As far as we are aware, ours is the first application of the Prugniel-Simien 
model to $N$-body halos. 

As in Merritt et al. (2005), we base our model evaluations
on {\it non}parametric representations of the $N$-body
density profiles.
Such representations are ``optimum'' in terms of their
bias-variance tradeoff, but are also notable for their flexibility.
Not only do they constitute (i) ``stand-alone,'' smooth and continuous
representations of the density and its slope:
they are also well suited to (ii) inferring best-fit values
for the fitting parameters of parametric functions, and
(iii) comparing the goodness-of-fit of different parametric
models, via the relative values of the integrated square
error or a similar statistic.
The more standard technique of computing binned
densities is suitable (though inferior) for (ii) and (iii) but
not for (i), since the density is given only at a discrete 
set of points and the derivatives poorly defined; 
while techniques like Sarazin's (1980) maximum-likelihood 
algorithm provide a (perhaps) more direct route to (ii) but are not 
appropriate for (i) or (iii).
Recently, nonparametric function estimation methods have
been applied to many other problems in astrophysics,
including reconstruction of the CMB fluctuation spectrum
(Miller et al. 2002), dynamics of dwarf galaxies (Wang et al. 2005),
and reconstruction of dark matter distributions via gravitational
lensing (Abdelsalam, Saha \& Williams 1998).
Application of nonparametric methods to the halo density profile 
problem is perhaps overdue, especially given the importance
of determining the inner density slope (Diemand et al. 2005).

In \S2 we introduce the data sets to be analyzed.  
These consist of $N$-body simulations of ten $\Lambda$CDM halos and 
two halos formed by monolithic (nearly spherical)
collapse.
(Moore et al. 1999 have discussed the similarity between the end state 
of cold collapse simulations and hierarchical CDM models.)
In \S3 we present the nonparametric method used to construct the
density profiles and their logarithmic slopes.
\S4 presents four, 3-parameter models,
and \S5 reports how well these empirical models perform.
Our findings are summarized in Section~\ref{SecSum}. 

In Paper II of this series (Graham et al.\ 2006a), we explore the 
Einasto and Prugniel-Simien models in more detail.  
Specifically, we explore
the logarithmic slope of these models and compare the results with
observations of real galaxies.  
We also present the models' circular
velocity profiles and their $\rho/\sigma^3$ profiles.
Helpful expressions for the concentration and assorted scale radii:
$r_s, r_{-2}, r_{\rm e}, R_{\rm e}, r_{\rm vir}$, and $r_{\rm max}$
--- the radius where the circular velocity profile has its maximum
value --- are also derived. 
Because the Prugniel-Simien model yields
the same parameters as those coming from S\'ersic-model fits,
we are able to show in Paper III
(Graham et al.\ 2006b) 
the location of our dark matter
halos on the Kormendy diagram ($\mu_{\rm e}$ vs.\ $\log R_{\rm e}$), 
along with real galaxies. 
We additionally show in Paper III the
location of our dark matter halos and real galaxies and clusters in a
new $\log(\rho_{\rm e})-\log(R_{\rm e})$ diagram. 

\section{Data: Dark matter halos\label{SecData}}

We use a sample of relaxed, dark matter halos from Diemand, Moore, \&
Stadel (2004a,b).  Details about the simulations, convergence tests,
and an estimate of the converged scales can be found in those papers.
Briefly, the sample consists of six, cluster-sized halos (models: A09,
B09, C09, D12, E09, and F09) resolved with 5 to 25 million particles
within the virial radius, and four, galaxy-sized halos (models: G00,
G01, G02, and G03) resolved with 2 to 4 million particles.
The innermost resolved radii are 0.3\% to 0.8\% of the virial radius,
$r_{\rm vir}$.  The outermost data point is roughly at the virial radius,
which is defined in such a way that the mean density within $r_{\rm 
vir}$ is $178 \Omega_M^{0.45} \rho_{\rm crit} = 98.4 \rho_{\rm crit}$
(Eke, Cole, \& Frenk 1996) using $\Omega_m = 0.268$ (Spergel et al.\ 2003). 
The virial 
radius thus encloses an overdensity which is 368 times denser than the
mean matter density.
We adopted the same estimates of the halo centers as in the
Diemand et al. papers; these were computed using SKID (Stadel 2001),
a kernel-based routine.

In an effort to study the similarities between cold, collisionless
collapse halos and CDM halos, we performed two additional simulations.
We distributed $10^7$ particles with an initial density profile
$\rho(r)\propto r^{-1}$, within a unit radius sphere with total mass 1
(M11) and 0.1 (M35).  The particles have zero kinetic energy and the
gravitational softening was set to 0.001.  
Each system collapsed and underwent a radial-orbit instability 
(Merritt \& Aguilar 1985)
which resulted in a virialized, triaxial/prolate structure. 
The lower mass halo, M35, collapsed less violently over a longer 
period of time.

\section{Nonparametric estimation of density profiles 
and their derivatives}\label{Dave}

Density profiles of $N$-body halos are commonly constructed by
counting particles in bins.
While a binned histogram is a bona-fide, non-parametric estimate of the
``true'' density profile, it has many undesirable
properties, e.g. it is discontinuous, and it depends sensitively on
the chosen size and location of the bins
(see, e.g., Stepanas \& Saha 1995).
A better approach is to view the particle positions as a random sample
drawn from some unknown, smooth density $\rho({\bf r})$, and to use
techniques from nonparametric function estimation to construct an
estimate $\hat\rho$ of $\rho$ (e.g., Scott 1992).  
In the limit that the ``sample size'' $N$ tends to infinity, 
such an estimate exactly reproduce the density function from which the data
were drawn, as well as many properties of that function, e.g.\ its
derivatives (Silverman 1986).

We used a kernel-based algorithm for estimating $\rho(r)$,
similar to the algorithms described in Merritt \& Tremblay (1994)
and Merritt (1996).
The starting point is an estimate of the 3D density
obtained by replacing each particle at position ${\bf r}_i$ 
by a kernel of width $h_i$, and summing the kernel densities:
\begin{equation}
\hat\rho({\bf r}) = \sum_{i=1}^N {m_i\over h_i^3} K\left[{1\over h_i}
\left|{\bf r} - {\bf r}_i\right|\right].
\label{eq:hatrho}
\end{equation}
Here $m_i$ is the mass associated with the $i$th particle
and $K$ is a normalized kernel function,
i.e.\ a density function with unit volume.
We adopted the Gaussian kernel,
\begin{equation}
K(y) = {1\over(2\pi)^{3/2}}{\rm e}^{-y^2/2}.
\end{equation}

The density estimate of equation~(\ref{eq:hatrho}) has no
imposed symmetries.
We now suppose that $\rho({\bf r}) = \rho(r)$, i.e.\ that
the underlying density is spherically symmetric about the
origin.
In order that the density estimate have this property,
we assume that
each particle is smeared uniformly around the surface
of the sphere whose radius is $r_i$.  
The spherically-symmetrized density estimate is
\begin{mathletters}
\begin{eqnarray}
\hat\rho(r) &=& \sum_{i=1}^N {m_i\over h_i^3} {1\over 4\pi} \int d\phi \int
d\theta\ \sin\theta\ K\left({{\rm d}\over h_i}\right), \\
{\rm d}^2 &=& \left|{\bf r} - {\bf r}_i\right|^2 \\
&=& r_i^2 + r^2 - 2rr_i\cos\theta
\end{eqnarray}
\end{mathletters}
\noindent
where $\theta$ is defined (arbitrarily) from the ${\bf r}_i$-axis.
This may be expressed in terms of the angle-averaged kernel $\tilde{K}$,
\begin{mathletters}
\begin{eqnarray}
\tilde{K}(r,r_i,h_i) &\equiv& {1\over 4\pi} \int_{-\pi/2}^{\pi/2} d\phi \nonumber\\
& & \hskip-80pt \times 
\int_0^{2\pi} d\theta\ \sin\theta\ K 
\left({h_i}^{-1}\sqrt{r_i^2 + r^2 - 2rr_i\cos\theta}\right)\\
&& \hskip-70pt = {1\over 2} \int_{-1}^1 d\mu\ K\left({h_i}^{-1}\sqrt{r_i^2 + r^2 -
2rr_i\mu}\right),
\end{eqnarray}
\end{mathletters}
\noindent
as
\begin{equation}
\hat\rho(r) = \sum_{i=1}^N {m_i\over h_i^3} \tilde{K}(r,r_i,h_i).
\label{eq:rhohat}
\end{equation}
Substituting for the Gaussian kernel, we find
\begin{eqnarray}
\tilde{K}(r,r_i,h_i) & = & {1\over (2\pi)^{3/2}} \left({r r_i\over
h_i^2}\right)^{-1} \nonumber \\
 & & \hskip-60pt \times \exp\left[-(r_i^2+r^2)/2h_i^2\right] \sinh(rr_i/h_i^2).
\end{eqnarray}
A computationally preferable form is
\begin{eqnarray}
\tilde{K} & = & {1\over 2(2\pi)^{3/2}} \left({r r_i\over h_i^2}\right)^{-1} \nonumber \\
& & \hskip-60pt \times
\left\{ 
\exp\left[-(r_i-r)^2/2h_i^2\right] - 
\exp\left[-(r_i+r)^2/2h_i^2\right]
\right\} .
\label{eq:ktilde}
\end{eqnarray}
Equations~(\ref{eq:rhohat}) and (\ref{eq:ktilde}) define
the density estimate.
Typically, one sets up a grid in radius and evaluates
$\hat\rho(r)$ discretely on the grid.
However we stress that the density estimate itself is
a continuous function and is defined independently of 
any grid.

Given a sample of $N$ positions and particle masses drawn
randomly from some (unknown) $\rho(r)$, the goal is to construct
an estimate $\hat\rho(r)$ that is as close as
possible, in some sense, to $\rho(r)$.
In the scheme just described, one has the freedom
to adjust the $N$ kernel widths $h_i$ in order to achieve
this.
In general, if the $h_i$ are too small, the density estimate
will be ``noisy,'' i.e.\ $\hat\rho(r)$ will exhibit
a large {\it variance} with respect to the true density;
while if the $h_i$ are too large, the density estimate
will be over-smoothed, i.e.\ there will be a large
{\it bias}.
(Of course the same is true for binned histograms,
although in general the bias-variance tradeoff for
histograms is less good than for kernel estimates.)
If the true $\rho(r)$ were known {\it a priori}, one
could adjust the $h_i$ so as to minimize (say) the
mean square deviation between $\rho(r)$ and $\hat\rho(r)$.
Since $\rho(r)$ is not known {\it a priori} for
our halos, some algorithm must be adopted for choosing
the $h_i$.
We followed the standard practice (e.g. Silverman 1986, p.101)
of varying the $h_i$ as a power of the local density:
\begin{equation}\label{hrhog}
h_i = h_0 \left[\hat\rho_{\rm pilot}(r_i)/g\right]^{-\alpha},
\end{equation}
where $\hat\rho_{\rm pilot}(r)$
is a ``pilot'' estimate of $\rho(r)$,
and $g$ is the geometric mean of the pilot densities 
at the $r_i$.
Since the pilot estimate is used only for assigning the
$h_i$, it need not be differentiable,
and we constructed it using a nearest-neighbor scheme.

\begin{figure}
\includegraphics[scale=0.41,angle=0]{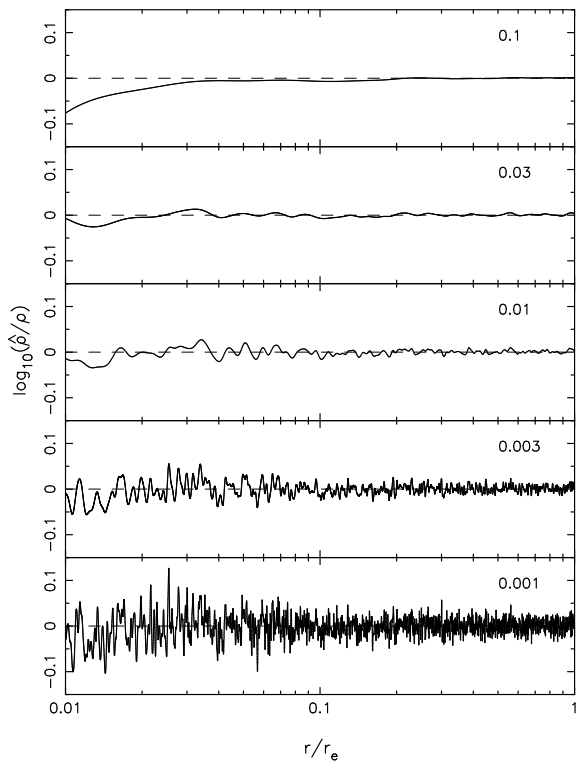}
\caption{\label{fig:tests1}
Nonparametric, bias-variance 
tradeoff in the estimation of $\rho(r)$ using a
single sample of $10^6$ radii generated from a halo having an 
Einasto $r^{1/n}$ density profile with $n=5$ (see Section~\ref{SecSer}). 
From top to bottom, $h_0=(0.1,0.03,0.01,0.003,0.001)\,r_{\rm e}$; 
all estimates used $\alpha=0.3$
(see equations~\ref{eq:rhohat}, \ref{eq:ktilde}, and \ref{hrhog}).
}
\end{figure}

The final density estimate $\hat\rho(r)$ is then 
a function of two quantities: $h_0$ and $\alpha$.
Figure~\ref{fig:tests1} illustrates the dependence of $\hat\rho(r)$
on $h_0$ when the kernel algorithm is applied to a 
random sample of $10^6$ equal-mass particles
generated from an Einasto density profile with $n=5$, 
corresponding to typical values observed in Merritt et al.\ (2005). 
Each of the density profile estimates of Figure~\ref{fig:tests1} used 
$\alpha=0.3$.
As expected, for small $h_0$, the estimate of $\rho(r)$
is noisy, but faithful in an average way to the true
profile; while for large $h_0$,  $\rho(r)$ is
a smooth function but is biased 
at small radii due to the averaging effect of the kernel.
For $\alpha=0.3$, 
The ``optimum'' $h_0$ for this sample is $\sim 0.05\,r_{\rm e}$, 
where $r_{\rm e}$ is the half-mass radius coming from the Einasto
model (see Section~\ref{SecSer}).

In what follows, we will compare the nonparametric estimates
$\hat\rho(r)$ derived from the $N$-body models with various parametric
fitting functions, in order to find the best-fitting parameters of the
latter by minimizing the rms residuals between the two profiles.  For
this purpose, any of the density estimates in Figure~\ref{fig:tests1}
would yield similar results, excepting perhaps the density estimate in
the uppermost panel which is clearly biased at small radii.  In
addition, we will also wish to characterize the rms value of the
deviation between the ``true'' profile and the best-fitting parametric
models.  Here it is useful for the kernel widths to be chosen such
that the residuals are dominated by the systematic differences between
the parametric and nonparametric profiles, and not by noise in
$\hat\rho(r)$ resulting from overly-small kernels.  We verified that
this condition was easily satisfied for all of the $N$-body models
analyzed here: there was always found to be a wide range of
$(h_0,\alpha)$ values such that the residuals between $\hat\rho(r)$
and the parametric function were nearly constant with varying radial 
coordinate.  This is a 
consequence of the large particle numbers ($>10^6$) in the $N$-body
models, which imply a low variance even for small $h_0$. 

\begin{figure}
\includegraphics[scale=0.41,angle=0]{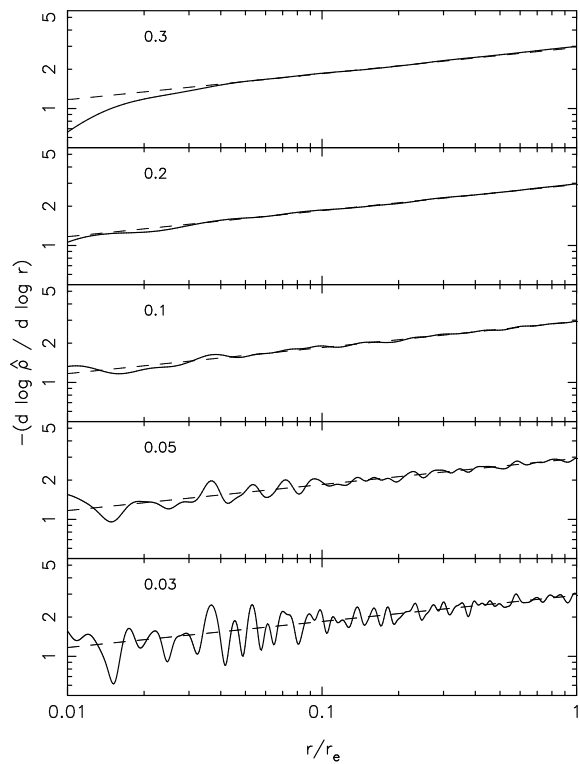}
\caption{\label{fig:tests2}
Five estimates of the logarithmic slope of an Einasto $r^{1/n}$ 
halo, derived via differentiation of $\hat\rho(r)$.
The same sample of $10^6$ radii was used as
in Figure~\ref{fig:tests1}.
From top to bottom, $h_0=(0.3,0.2,0.1,0.05,0.03)\,r_{\rm e}$; 
each estimate used $\alpha=0.4$ (see equation~\ref{hrhog}).
Dashed lines show the true slope.
}
\end{figure}

As discussed above, quantities like the derivative
of the density can also be computed directly from $\hat\rho(r)$.
Figure~\ref{fig:tests2} shows nonparametric estimates of the slope,
$d\log\rho/d\log r$,
for the same $10^6$ particle data set as in Figure~\ref{fig:tests1}.
We computed derivatives simply by numerically differentiating
$\hat\rho(r)$; alternatively, we could have differentiated
equation~(\ref{eq:ktilde}). 
Figure~\ref{fig:tests2} shows that as $h_0$ is increased, the variance in
the estimated slope drops, and for $h_0\approx 0.2\,r_{\rm e}$ 
the estimate is very close to the true function.
We note that the optimal choice of $h_0$ when estimating
derivatives is larger than when estimating $\rho(r)$
($\sim 0.2\,r_{\rm e}$ vs.\ $\sim 0.05\,r_{\rm e}$); 
this is a well-known consequence of the increase in ``noise''
associated with differentiation.
Figure~\ref{fig:tests2} also illustrates the important point that there is no
need to impose an additional level of smoothing
when computing the density derivatives (as was done, e.g., in
Reed et al.\ 2005); it is sufficient to increase $h_0$.

\subsection{Application to the $N$-body halos}

\begin{figure*}
\includegraphics[scale=0.75,angle=0]{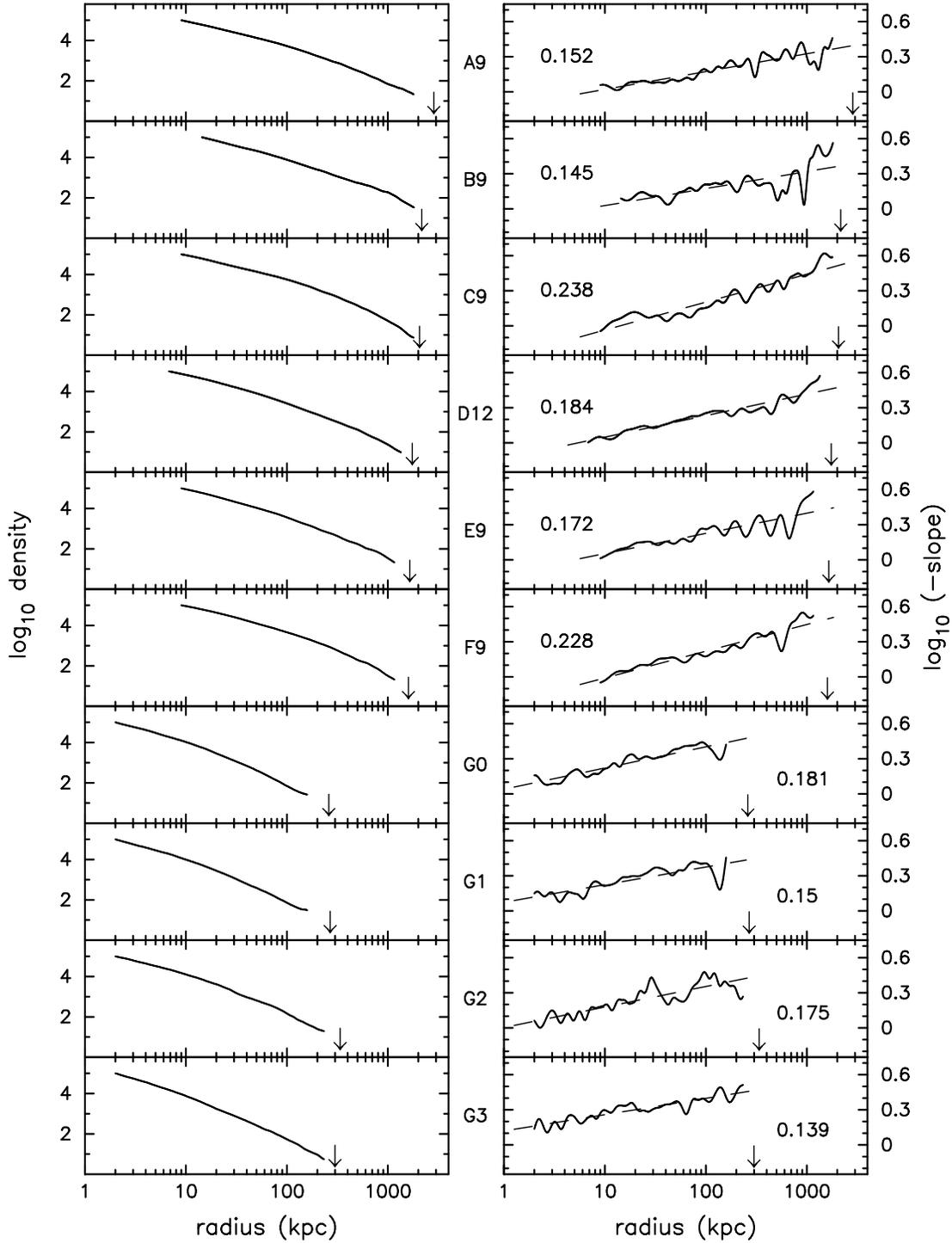}
\caption{\label{fig:profs1}
Nonparametric estimates of the density $\rho(r)$ (left panel) and
the slope $d\log\rho/d\log r$ (right panel) for the ten $N$-body halos
of Table~\ref{Table1}.
The virial radius $r_{vir}$ is marked with an arrow.
Dashed lines in the right hand panels are linear fits
of $\log(-d\log\rho/d\log r)$ to $\log r$;
regression coefficients are also given.
}
\end{figure*}

\begin{figure*}
\includegraphics[scale=0.84,angle=0]{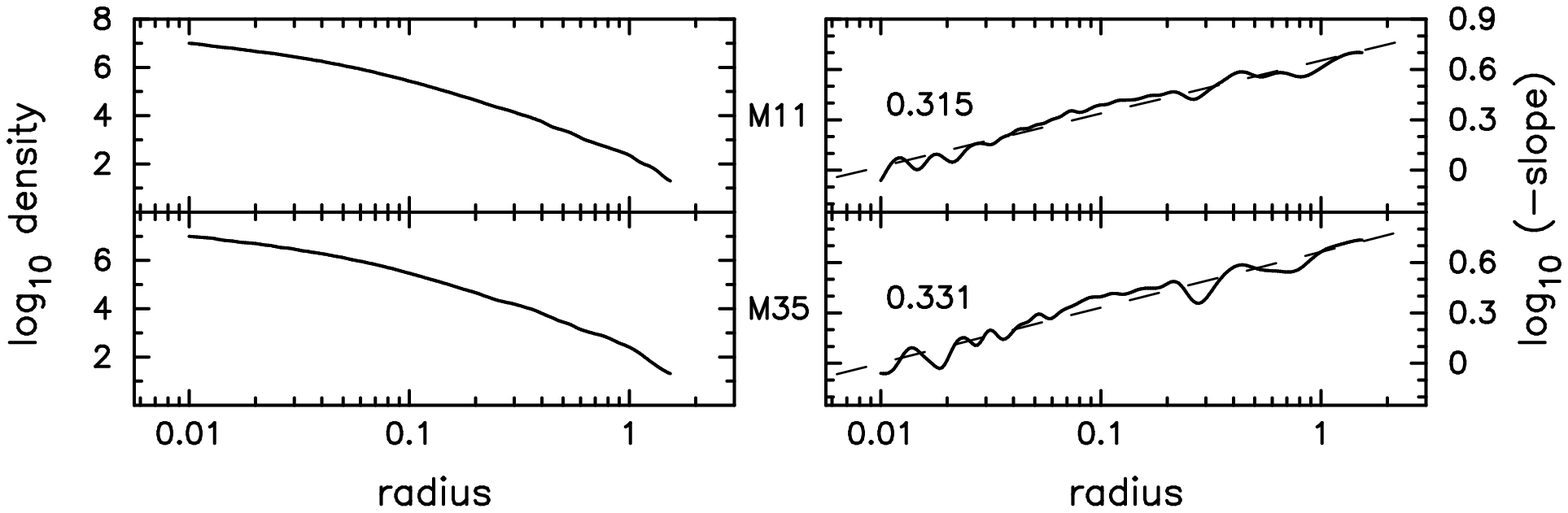}
\caption{\label{fig:profs2}
Nonparametric estimates of $\rho(r)$ (left panel) and
$d\log\rho/d\log r$ (right panel) for the two
``collapse'' models.
Dashed lines in the right hand panels are linear fits
of $\log(-d\log\rho/d\log r)$ to $\log r$.
}
\end{figure*}

Figure~\ref{fig:profs1} shows, using 
$\alpha = 0.3$ and $h_0 = 0.05\,r_{\rm e}$ (left panel) and 
$\alpha = 0.4$ and $h_0 = 0.05\,r_{\rm e}$ (right panel)\footnote{We 
have purposely used a small value of $h_0$ to 
avoid any possibility of biasing the slope estimates.}, 
the nonparametric estimates of $\rho(r)$ (left panel) 
and $\gamma(r)\equiv d\log\rho/d\log r$ (right panel) 
for the ten $N$-body halos. 
Figure~\ref{fig:profs2} shows the same quantities for the two data sets
generated from cold collapses.  
We stress that these plots -- especially, the derivative
plots -- could not have been made from
tables of binned particle numbers.
For most profiles, the slope is a rather continuous function of radius
and does not appear to reach any obvious, asymptotic, central value by
$\sim 0.01 r_{\rm vir}$.  Instead, $\hat\gamma(r)$ varies
approximately as a power of $r$, i.e.\ $\log\hat\gamma$ vs.\ $\log r$
is approximately a straight line.  Accordingly, we have fitted
straight lines, via a least-squares minimization, to the logarithmic
profile slopes in the right-hand panels of Figures~\ref{fig:profs1}
and \ref{fig:profs2}.  The regression coefficients, i.e.\ slopes, are
inset in each panel.  (These slope estimates should be seen as
indicative only; they are superseded by the model fits discussed
below.) In passing we note that such a power-law
dependence of $\gamma$ on $r$ is characteristic of the Einasto  
model, with the logarithmic slope equal to the exponent $1/n$.
%
Noise and probable (small) deviations from a perfect Einasto $r^{1/n}$
model are expected to produce slightly different exponents
when we fit the density profiles in the following Section with 
Einasto's $r^{1/n}$ model and a number of other empirical functions. 

The slope at the innermost resolved radius is always close
to $-1$, which is also the slope at $r=0$ in
the NFW model. 
However there is no indication in Figure 3 
that $\hat\gamma(r)$ is flattening
at small radii, i.e., it is natural to conclude that $N$-body
simulations of higher resolution would exhibit smaller inner
slopes.
On average, the slope at $r_{\rm vir}$ is around $-3$, but there are
large fluctuations and some halos reach a value of $-4$, as previously noted in
Diemand et al.\ (2004b).  The reason for these fluctuation may be
because the outer parts are dynamically very young (i.e.\ measured in
local dynamical times) and they have only partially completed the
violent relaxation into a stable, stationary equilibrium configuration.
We are not able to say with any confidence what the slopes do beyond $r_{\rm vir}$.


\section{Empirical models}\label{SecMod}

In this section, we present 
four parametric density models, each having three independent parameters:
two ``scaling'' parameters and one ``shape'' parameter.
We measured the quality of each parametric model's fit to the 
nonparametric $\hat\rho(r)$'s using a standard metric, the integrated 
square deviation,
\begin{equation}
\label{EqISE}
\int d\left(\log r\right) 
\left[\log\hat\rho(r) - \log\rho_{\rm param}(r)\right]^2
\end{equation}
where $\rho_{\rm param}$ is understood to depend on the
various fitting parameters as well as on $r$.
Equation~(\ref{EqISE}) is identical in form to the
Cram\'er-von Mises statistic (e.g. Cox \& Hinkley 1974, eq. 6),
an alternative to the Kolmogorov-Smirnov statistic for
comparing two (cumulative) distribution functions.

We chose to evaluate this integral by discrete summation
on a grid  spaced uniformly in $\log r$; our
measure of goodness-of-fit (which was also the quantity
that was minimized in determing the best-fit parameters) was
\begin{mathletters}
\begin{eqnarray}\label{EqChi}
\Delta^2 &\equiv& {\sum_{j=1}^m\delta_j^2\over m-3}, \\
\delta_j &=& \log_{10}[\hat\rho(r_j)/\rho_{\rm param}(r_j)]
\end{eqnarray}
\end{mathletters}
with $m=300$.  
With such a large value of $m$ the results obtained
by minimizing (\ref{EqChi}) and (\ref{EqISE}) are indistinguishable.
We note that the quantity $\Delta^2$ in equation~(\ref{EqChi})
is reminiscent of the standard $\chi^2$,
but the resemblance is superficial.
For instance, $\Delta^2$ as defined here is independent 
of $m$ in the large-$m$ limit 
(and our choice of $m=300$ puts us effectively in this limit).
Furthermore there is no binning involved in the computation
of $\Delta^2$; the grid is simply a numerical device used in
the computation of (\ref{EqISE}).

\subsection{Double power-law models}\label{GenNFW}

Hernquist (1990, his equation~43) presented a 5-parameter
generalization of Jaffe's (1983) double power-law model.
Sometimes referred to as the ($\alpha, \beta, \gamma$) model, it can be
written as
\begin{equation} \label{Eq13gam}
\rho(r)={\rho_s}2^{(\beta-\gamma)/\alpha}\left(\frac{r}{r_s}\right)^{-\gamma}
\left[1+\left( \frac{r}{r_s} \right)^{\alpha}\right]^{(\gamma-\beta)/\alpha}, 
\end{equation}
where $\rho_s$ is the density at the scale radius, $r_s$, which marks
the center of the transition region between the inner and outer
power-laws having slopes of $-\gamma$ and $-\beta$, respectively.  The
parameter $\alpha$ controls the sharpness of the transition (see Zhao
et al.\ 1996; Kravtsov et al.\ 1998; 
and equations~(37) and (40b) in Dehnen \& McLaughlin 2005).
%
%
Setting ($\alpha, \beta, \gamma$)=(1, 3, 1) yields the NFW model,
while (1.5, 3, 1.5) gives the model in Moore et al.\ (1999).
Other combinations have been used, for example, (1,3,1.5) was applied
in Jing \& Suto (2000) and (1,2.5,1) was used by Rasia, Tormen, \&
Moscardini (2004).

In fitting dark matter halos, Klypin et al.\ (2001, their figure~8)
have noted a certain degree of degeneracy when all of the 5 paramaters
are allowed to vary.  Graham et al.\ (2003, their figures~3 and 4) have
also observed the parameters of this empirical model to be highly
unstable when applied to (light) profiles having a continuously
changing logarithmic slope.  Under such circumstances, the parameters
can be a strong function of the fitted radial extent, rather than
reflecting the intrinsic physical properties of the profile under
study.  This was found to be the case when applied to the dark matter
halos under study here.  We have therefore chosen to constrain two of
the model parameters, holding $\alpha$ fixed at 1 and $\beta$ fixed at 3.

In recent years, as the resolution in $N$-body simulations has
improved, Moore and collaborators have found that the innermost
(resolved) logarithmic slope of dark matter halos has a range of
values which are typically shallower than $-1.5$: recently obtaining a
mean value ($\pm$ standard deviation) equal to $-1.26 \pm 0.17$ at 1\%
of the virial radius (Diemand, Moore, \& Stadel 2004b).
At the same time, Navarro et al.\ (2004) report that the NFW model
underestimates the density over the inner regions of most of their
halos, which have innermost resolved slopes ranging from $-1.6$ to $-0.95$
(their Figure~3).
A model with an outer slope of $-3$ and an inner slope of $-\gamma$ 
might therefore be more appropriate.
Such a model has been used before and can be written as 
\begin{equation}
\rho(r) = \frac{2^{3-\gamma}\rho_s}{(r/r_s)^{\gamma}(1+r/r_s)^{3-\gamma}}. 
\label{EqNFW}
\end{equation}
The total mass of  this model is infinite however.

\begin{figure}
\includegraphics[scale=0.40,angle=0]{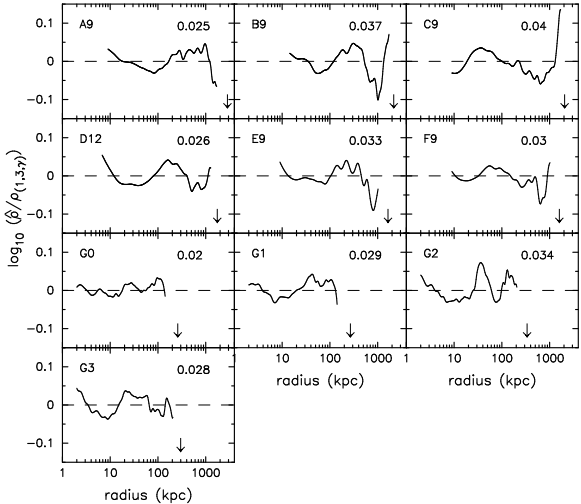}
\caption{
  Residual profiles from 
application of the 3-parameter (1, 3, $\gamma$) model (equation~\ref{EqNFW}) 
to our ten, $N$-body density profiles.  The virial radius is 
marked with an arrow, and the rms residual (equation~\ref{EqChi})
is inset with the residual profiles. 
}
\label{FigNFW}
\end{figure}

\begin{figure}
\includegraphics[scale=0.57,angle=0]{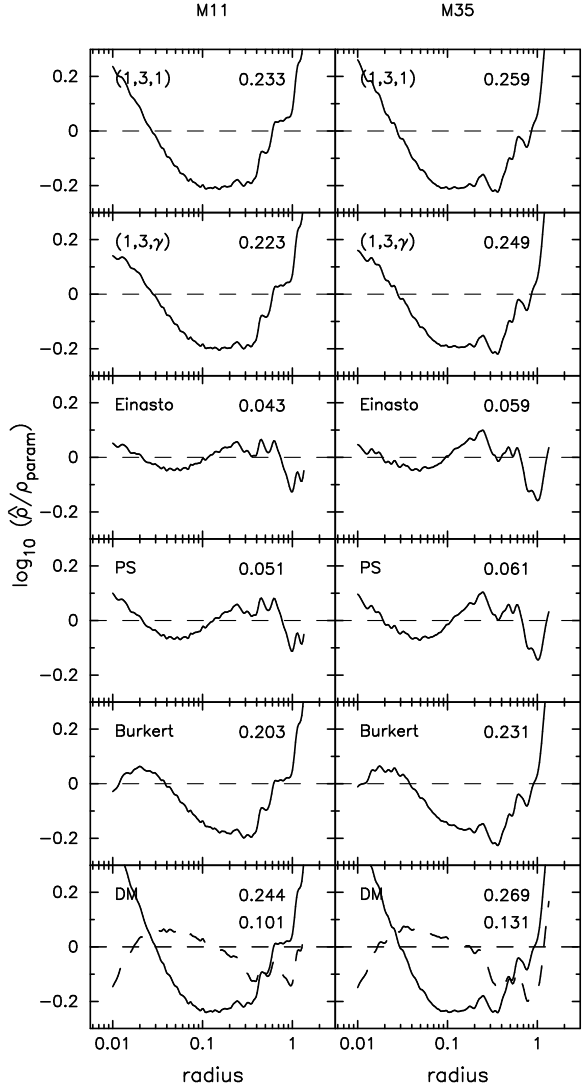}
\caption{
  Residual profiles from the application of seven different parametric 
models (see Section~\ref{SecMod}) 
to our ``cold collapse'' density halos, M11 and M35. 
Einasto's model, which has the same functional form as S\'ersic's
model, is labelled `Sersic' in this Figure. 
In the lower panel, the solid curve corresponds to the 2-parameter
model from Dehnen \& McLaughlin (2005), and the dashed curve corresponds
to their 3-parameter model. 
The rms residual (equation~\ref{EqChi}) is inset in each figure. 
}
\label{Figcold}
\end{figure}

We have applied the above (1, 3, $\gamma$) model to our dark matter
density profiles, the results of which are shown in
Figure~\ref{FigNFW} for the $N$-body halos, and in the upper panel of
Figure~\ref{Figcold} for the cold collapse models.  
The rms scatter, $\Delta$, is inset in each figure and 
additionally reported in Table~\ref{Table1}.

\begin{deluxetable}{lcccc|cccc|cccc}
\tablewidth{0pt}
\tabletypesize{\footnotesize}
\tablecaption{Three-Parameter Models}
\tablehead{ 
\colhead{Halo} &  
\colhead{$r_s$}       & \colhead{$\log \rho_s$} & \colhead{$\gamma$}       & \colhead{$\Delta$} &
\colhead{$r_{\rm e}$} & \colhead{$\log \rho_e$} & \colhead{$n_{\rm Ein}$}  & \colhead{$\Delta$} &
\colhead{$R_{\rm e}$} & \colhead{$\log \rho^{\prime}$} & \colhead{$n_{\rm PS}$} & \colhead{$\Delta$} \\
\colhead{Id.} &  
\colhead{kpc} & \colhead{$M_{\sun}$ pc$^{-3}$}  & \colhead{} & \colhead{dex} &
\colhead{kpc} & \colhead{$M_{\sun}$ pc$^{-3}$}  & \colhead{} & \colhead{dex} &
\colhead{kpc} & \colhead{$M_{\sun}$ pc$^{-3}$}  & \colhead{} & \colhead{dex}  
}
\startdata
\multicolumn{1}{c}{ }
 & \multicolumn{4}{c}{(1,3,$\gamma$)} & \multicolumn{4}{c}{Einasto $r^{1/n}$}
 & \multicolumn{4}{c}{Prugniel-Simien} \\
 \multicolumn{13}{c}{Cluster-sized halos} \\
 A09 &  626.9  & $-3.87$ & 1.174 & 0.025  &   5962.  & $-6.29$ & 6.007 & {\bf 0.015}  &  2329.  & $-2.73$ & 3.015 & 0.021  \\
 B09 & 1164.   & $-4.75$ & 1.304 & {\bf 0.037}  &  17380.  & $-7.66$ & 7.394 & 0.041  &  4730.  & $-3.34$ & 3.473 & 0.038  \\
 C09 &  241.8  & $-3.27$ & 0.896 & 0.040  &   1247.  & $-4.95$ & 3.870 & 0.030  &   738.9 & $-2.55$ & 2.192 & {\bf 0.016}  \\
 D12 &  356.1  & $-3.82$ & 1.251 & 0.026  &   2663.  & $-6.02$ & 5.939 & 0.020  &  1232.  & $-2.52$ & 3.147 & {\bf 0.019}  \\
 E09 &  382.5  & $-3.96$ & 1.265 & 0.033  &   2611.  & $-6.06$ & 5.801 & 0.032  &  1231.  & $-2.62$ & 3.096 & {\bf 0.030}  \\
 F09 &  233.9  & $-3.51$ & 1.012 & 0.030  &   1235.  & $-5.26$ & 4.280 & 0.025  &   697.3 & $-2.63$ & 2.400 & {\bf 0.017}  \\
 \multicolumn{13}{c}{Galaxy-sized halos} \\
 G00 &   27.96 & $-3.16$ & 1.163 & {\bf 0.020}  &    189.0 & $-5.22$ & 5.284 & 0.023  &   114.4 & $-2.02$ & 3.135 & 0.028  \\
 G01 &   35.34 & $-3.36$ & 1.275 & 0.029  &    252.6 & $-5.51$ & 5.873 & {\bf 0.028}  &   146.0 & $-2.01$ & 3.425 & 0.032  \\
 G02 &   53.82 & $-3.59$ & 1.229 & 0.034  &    391.4 & $-5.74$ & 5.725 & {\bf 0.031}  &   214.9 & $-2.34$ & 3.243 & 0.036  \\
 G03 &   54.11 & $-3.70$ & 1.593 & 0.028  &    405.6 & $-5.98$ & 7.791 & {\bf 0.023}  &   229.1 & $-1.47$ & 4.551 & 0.024  \\
 \multicolumn{13}{c}{Spherical collapse halos} \\
 M11 &  0.0175 &  2.66 & 0.006 & 0.223 &     0.244 &  0.27 & 3.426 & {\bf 0.043}  &   0.187 &  2.57 & 2.445 & 0.051  \\
 M35 &  0.0180 &  1.62 & 0.030 & 0.249 &     0.240 & $-0.70$ & 3.214 & {\bf 0.059}  &   0.185 &  1.47 & 2.301 & 0.061  \\
\enddata
\tablecomments{
Col.(1): Object Id. 
Col.(2)--(5) (1, 3, $\gamma$) model (equation~\ref{Eq13gam} and \ref{EqNFW}) 
scale radius $r_s$, scale density $\rho_s$, inner profile slope 
$\gamma$, and rms scatter of the fit.
Col.(6)--(9) Einasto $r^{1/n}$ model half-mass radius $r_{\rm e}$, associated density $\rho_{\rm e}$, 
profile shape $n_{\rm Ein}$, and rms scatter of the fit.
Col.(10)--(13) Prugniel-Simien model scale radius $R_{\rm e}$, 
scale density $\rho^{\prime}$ (the spatial density 
$\rho_{\rm e}$ at $r=R_{\rm e}$ is such that 
$\rho_{\rm e} = \rho^{\prime} {\rm e}^{-b}$), 
profile shape $n_{\rm PS}$, and rms scatter of the fit.
%
%
Note: The radius and density units do not apply to M11 and M35. 
For each halo, of the three models shown here 
the model having the lowest residual scatter is high-lighted in bold. 
\label{Table1}
}
\end{deluxetable}

\begin{deluxetable}{lcccc}
\tablewidth{0pt}
\tabletypesize{\footnotesize}
\tablecaption{Three-Parameter Models {\it (cont.)}}
\tablehead{ 
\colhead{Halo}  &  
\colhead{$r_s$} & \colhead{$\log \rho_s$} & \colhead{$\gamma^{\prime}$}   & \colhead{$\Delta$} \\
\colhead{Id.} &  
\colhead{kpc} & \colhead{$M_{\sun}$ pc$^{-3}$}  & \colhead{} & \colhead{dex} 
}
\startdata
\multicolumn{1}{c}{ }  & \multicolumn{4}{c}{Anisotropic Dehnen-McLaughlin (Eq.\ref{EqDMc3})} \\ 
 \multicolumn{5}{c}{Cluster-sized halos} \\
 A09 &  722.7  & $-2.21$ & 0.694 & {\bf 0.013}  \\
 B09 & 1722.   & $-3.30$ & 0.880 & 0.040  \\
 C09 &  207.0  & $-1.34$ & 0.241 & 0.047  \\
 D12 &  322.8  & $-1.95$ & 0.683 & 0.022  \\
 E09 &  330.4  & $-2.04$ & 0.669 & 0.034  \\
 F09 &  193.6  & $-1.56$ & 0.350 & 0.036  \\
 \multicolumn{5}{c}{Galaxy-sized halos} \\
 G00 &   20.89 & $-1.11$ & 0.422 & {\bf 0.017}  \\
 G01 &   25.88 & $-1.28$ & 0.568 & {\bf 0.023}  \\
 G02 &   43.05 & $-1.60$ & 0.581 & {\bf 0.027}  \\
 G03 &   30.20 & $-1.34$ & 0.849 &      0.024  \\
 \multicolumn{5}{c}{Spherical collapse halos} \\
 M11 &  0.025  & 4.23 & 0.00  & 0.179  \\
 M35 &  0.025  & 3.21 & 0.00  & 0.206  \\
\enddata
\tablecomments{
Col.(1): Object Id.  Col.(2)--(5) Dehnen-McLaughlin (their equation~46b) 
scale radius $r_s$, scale density $\rho_s$, inner profile slope 
$\gamma^{\prime}$, and rms scatter of the fit.
Note: The radius and density units do not apply to M11 and M35. 
When the rms scatter is lower than the value obtained with the other
3-parameter models, it is given in bold. 
\label{Table1b}
}
\end{deluxetable}

\subsubsection{Two-parameter models} \label{ProbPow}

\begin{figure}
\includegraphics[scale=0.40,angle=0]{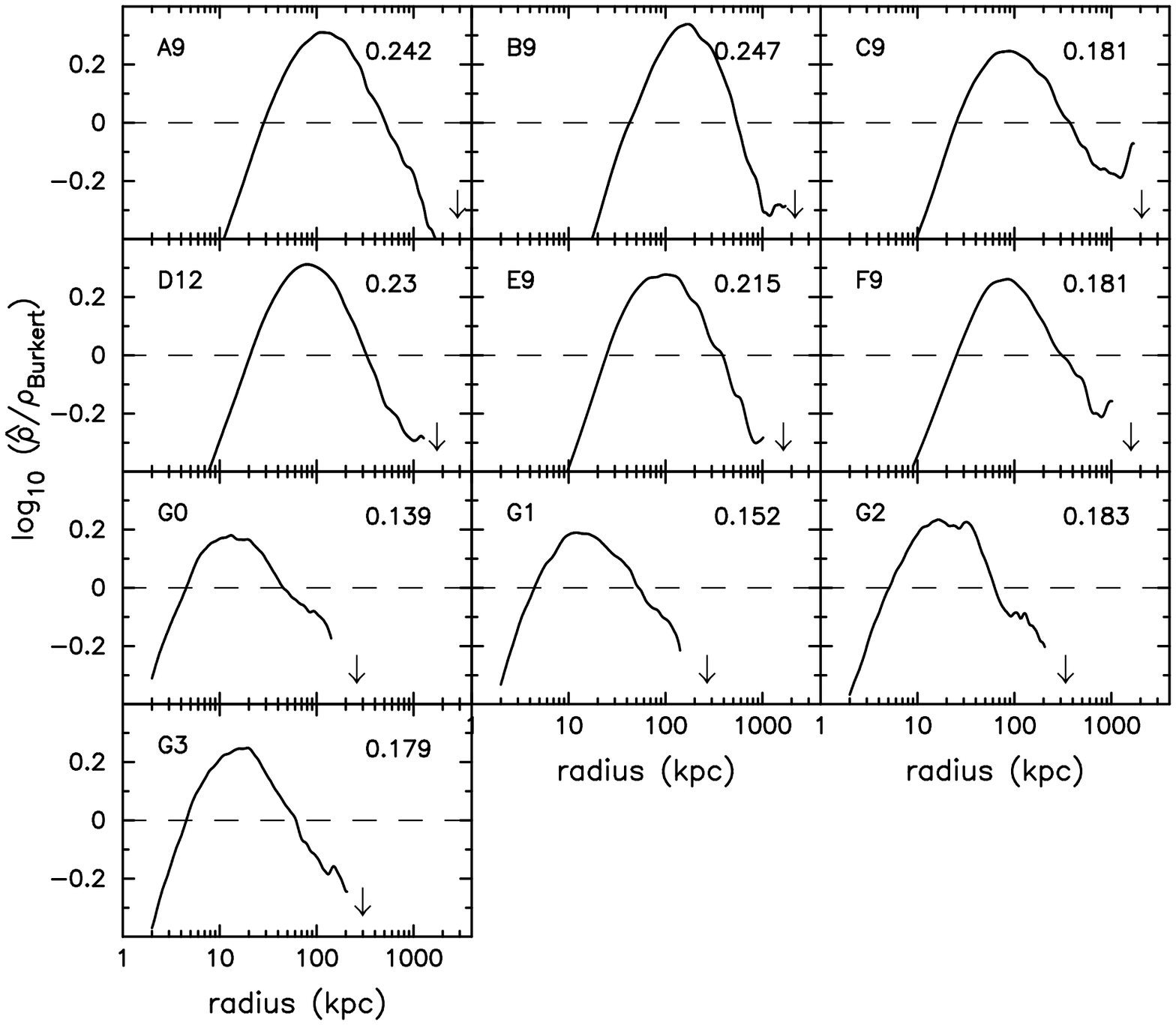}
\caption{
 Residual profiles from application of Burkert's 
2-parameter model (equation~\ref{EqBurk}) 
to our dark matter density profiles. 
%
}
\label{FigBurk}
\end{figure}

Recognizing that galaxies appear to have flat inner density profiles
(e.g., Flores \& Primack 1994; Moore 1994), Burkert (1995) cleverly 
introduced 
a density model having an inner slope of zero and an outer profile
that decayed as $r^{-3}$.  His model is given by the expression
\begin{equation}
\rho(r) = \frac{\rho_0 {r_s}^3}{(r+r_s)(r^2+{r_s}^2)}, 
\label{EqBurk}
\end{equation}
where $\rho_0$ is the central density and $r_s$ is a scale radius. 
Application of this model in Figure~\ref{FigBurk} reveals that, with only 
2 free parameters, it does not provide as good a fit to the simulated dark 
matter halos as the (1, 3, $\gamma$) model presented above.  The hump-shaped
residual profiles in Figure~\ref{FigBurk} signify the model's inability
to match the curvature of our density profiles.  
(It is important to point out that Burkert's model was introduced
to fit the observed rotation curves in low surface
brightness galaxies, after the contribution from the baryonic
component had been subtracted out, a task which it performs well.)

\begin{deluxetable}{lccc|ccc|ccc}
\tablewidth{0pt}
\tablecaption{Two-Parameter Models}
\tablehead{
\colhead{Halo} &
\colhead{$r_s$}       & \colhead{$\log \rho_0$} &  \colhead{$\Delta$} &
\colhead{$r_s$}       & \colhead{$\log \rho_s$} &  \colhead{$\Delta$} &
\colhead{$r_{\rm s}$} & \colhead{$\log \rho_s$} &  \colhead{$\Delta$} \\
\colhead{Id.} &
\colhead{kpc} & \colhead{$M_{\sun}$ pc$^{-3}$}  & \colhead{dex} &
\colhead{kpc} & \colhead{$M_{\sun}$ pc$^{-3}$}  & \colhead{dex} &
\colhead{kpc} & \colhead{$M_{\sun}$ pc$^{-3}$}  & \colhead{dex}
}
\startdata
\multicolumn{1}{c}{ }
 & \multicolumn{3}{c}{Burkert} 
 & \multicolumn{3}{c}{NFW} 
 & \multicolumn{3}{c}{Isotropic Dehnen-McLaughlin (equation~\ref{EqDMc2})} \\
       \multicolumn{10}{c}{Cluster-sized halos} \\
 A09  &  114.0 & $-1.65$ & 0.242 &    419.8 & $-3.50$ & 0.042 &    933.7 & $-2.43$ & {\bf 0.018} \\
 B09  &  145.2 & $-2.23$ & 0.247 &    527.2 & $-4.03$ & 0.068 &   1180.0 & $-2.97$ & {\bf 0.042} \\
 C09  &  96.16 & $-1.74$ & 0.181 &    284.4 & $-3.42$ & {\bf 0.042} &    554.3 & $-2.27$ & 0.091 \\
 D12  &  68.39 & $-1.62$ & 0.230 &    213.3 & $-3.34$ & 0.051 &    409.1 & $-2.17$ & {\bf 0.026} \\
 E09  &  77.09 & $-1.80$ & 0.215 &    227.0 & $-3.46$ & 0.053 &    428.2 & $-2.28$ & {\bf 0.037} \\
 F09  &  80.17 & $-1.85$ & 0.181 &    229.0 & $-3.49$ & {\bf 0.030} &    438.2 & $-2.32$ & 0.066 \\
        \multicolumn{10}{c}{Galaxy-sized halos} \\
 G00  &  10.12 & $-1.56$ & 0.139 &    22.23 & $-2.94$ & {\bf 0.024} &    34.43 & $-1.59$ & 0.037 \\
 G01  &  10.28 & $-1.54$ & 0.152 &    23.12 & $-2.95$ & 0.038 &    36.53 & $-1.61$ & {\bf 0.031} \\
 G02  &  14.06 & $-1.66$ & 0.183 &    36.39 & $-3.22$ & 0.044 &    63.06 & $-1.96$ & {\bf 0.035} \\
 G03  &  09.35 & $-1.32$ & 0.179 &    19.54 & $-2.68$ & 0.066 &    26.98 & $-1.23$ & {\bf 0.025} \\
   \multicolumn{10}{c}{Spherical collapse halos} \\
 M11  & 0.0261 &  3.01 & {\bf 0.203} &    0.0309 & 2.23 & 0.233 &    0.0234 & 4.31 & 0.244 \\
 M35  & 0.0265 &  1.98 & {\bf 0.231} &    0.0314 & 1.20 & 0.259 &    0.0236 & 3.29 & 0.269 \\
\enddata
\tablecomments{
Col.(1): Object Id. 
Col.(2)--(4) Burkert (1995) model scale radius $r_s$, 
central density $\rho_0$, and rms scatter of the fit 
(using $m-2$ in the denominator of equation~ref{EqChi}). 
Col.(5)--(7) NFW (1, 3, 1) model scale radius $r_s$,
scale density $\rho_0$, and rms scatter of the fit (using $m-2$). 
Col.(8)--(10) Dehnen-McLaughlin (2005, their equation~20b) model scale 
radius $r_{\rm s}$, associated density $\rho_{\rm s}$, and rms scatter
(using $m-2$). 
This model has an inner and outer, negative logarithmic 
slope of 7/9 $\approx$ 0.78 and 31/9 $\approx$ 3.44, respectively. 
Note: The above radius and density units do not apply to M11 and M35.
For each halo, the 2-parameter model with the lowest residual scatter is high-lighted in bold. 
}
\end{deluxetable}

\begin{figure}
\includegraphics[scale=0.40,angle=0]{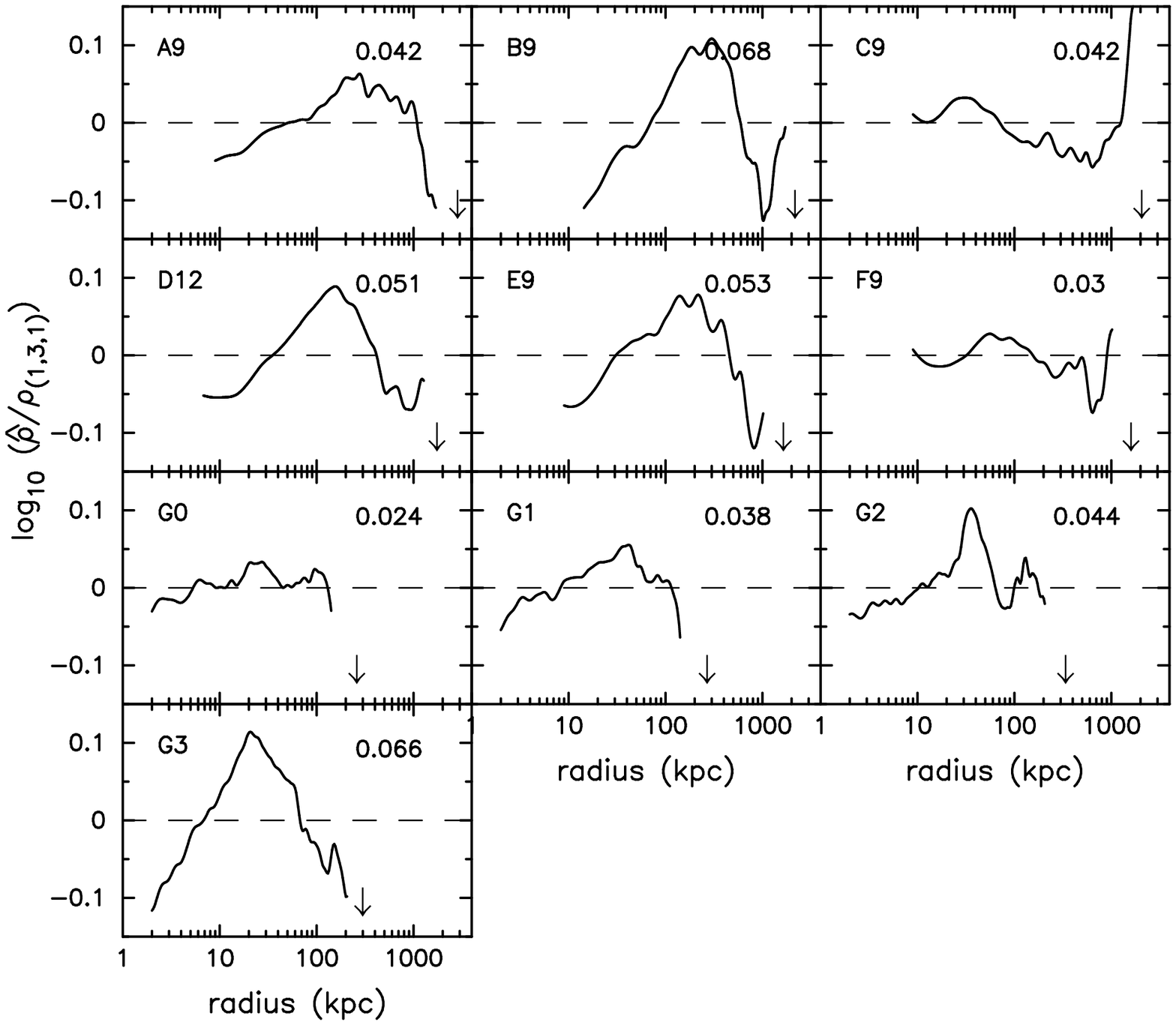}
\caption{
  Residual profiles from application of the 2-parameter 
NFW (1, 3, 1) model to our dark matter density profiles.  
}
\label{Fig131}
\end{figure}

As noted previously, the NFW ($\alpha, \beta, \gamma$)=(1, 3, 1) model
also has only two parameters: $\rho_s$ and $r_s$.  Because this model
is still often used, we apply it to our halos in Figure~\ref{Fig131}.
Comparison with Figure~\ref{FigNFW} reveals that the NFW model never
performs better than the (1, 3, $\gamma$) model; the residuals are
$\sim$50\% larger and sometimes twice as large. Importantly, the
large-scale curvature observed in many of the NFW resdiual profiles
(Figure~\ref{Fig131}) reveals that this model does not describe the
majority of the halos, and that the (1, 3, $\gamma$) model should be
preferred over the NFW model.  

\begin{figure}
\includegraphics[scale=0.40,angle=0]{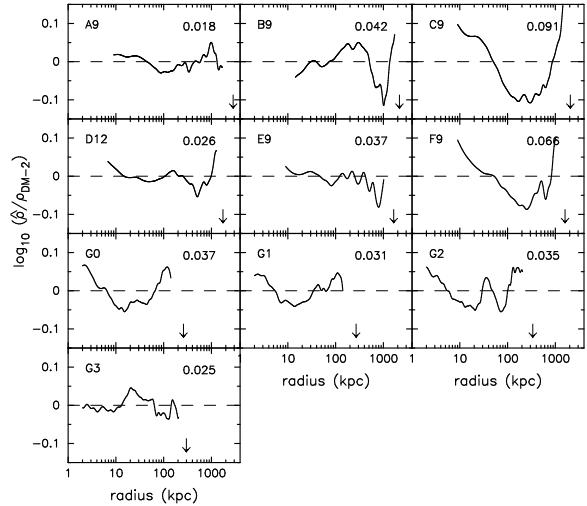}
\caption{
  Residual profiles from 
application of the 2-parameter (4/9, 31/9, 7/9) model 
(equation~\ref{EqDMc2}) 
from Dehnen \& McLaughlin (2005, their equation~20b) 
to our dark matter density profiles.  
}
\label{FigDMc}
\end{figure}

An alternative 2-parameter expression has recently been studied by
Dehnen \& McLaughlin (2005, their equation~(20b); see also Austin et
al.\ 2005).  It is a special case of a more general family of
models --- which we test next --- when the velocity ellipsoid at the
halo center is isotropic and $\rho/{\sigma_r}^3$ is a (special)
power law in radius, varying as $r^{-35/18}$.  This 2-parameter density 
model is an ($\alpha, \beta, \gamma) = (4/9, 31/9, 7/9$) model given
by
\begin{equation}
\rho(r) = \frac{2^6\rho_s}{(r/r_s)^{7/9}[1+(r/r_s)^{4/9}]^6}, 
\label{EqDMc2}
\end{equation}
and is applied in Figure~\ref{FigDMc}.  
It clearly provides a much better match to the dark matter halo density 
profiles in comparison with the previous 2-parameter model 
over the fitted radial range, but the rms scatter reveals that it 
does not perform as well as the (1, 3, $\gamma$) model,
nor can it describe the `spherical collapse' halos (Figure~\ref{Figcold}). 
We therefore, in the following subsection, test the more general
3-parameter model given in Dehnen \& McLaughlin (2005). 

\subsubsection{Dehnen-McLaughlin's anisotropic 3-parameter model}

Dehnen \& McLaughlin (2005, their equation~46b) present a
theoretically-motivated, 3-parameter model such that [$\alpha, \beta,
\gamma] = [ 2(2-\beta_0)/9, (31-2\beta_0)/9, (7+10\beta_0)/9 ]$, and
the term $\beta_0$ reflects the central ($r=0$) anisotropy --- a
measure of the tangential to radial velocity dispersion\footnote{Note:
the quantities $\beta$ and $\beta_0$ are not as related as their
notation suggests.  The former is the outermost, negative logarithmic
slope of the density profile while the latter is the velocity
anisotropy parameter at $r=0$.}.
Setting $\gamma^{\prime} = (7+10\beta_0)/9$, we have
$[\alpha, \beta, \gamma] = 
[ (3-\gamma^{\prime})/5, (18-\gamma^{\prime})/5, \gamma^{\prime} ]$, and 
their density model can be written as 
\begin{equation}
\rho(r) = \frac{2^6\rho_s}{(r/r_s)^{\gamma^{\prime}}[1+(r/r_s)^{(3-\gamma^{\prime})/5}]^6}, 
\label{EqDMc3}
\end{equation}

\begin{figure}
\includegraphics[scale=0.40]{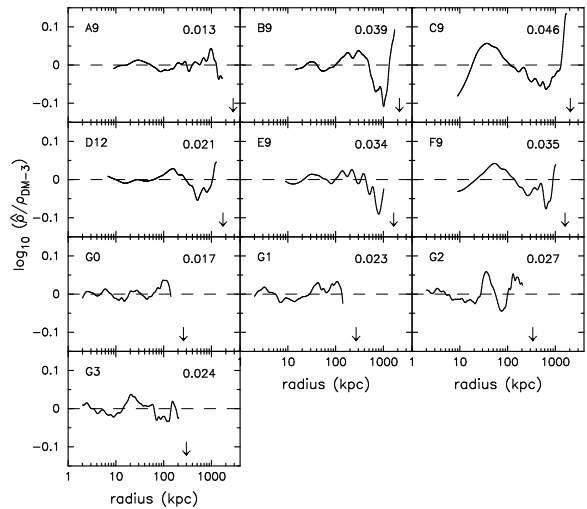}
\caption{
  Residual profiles from application of the 3-parameter 
$[(3-\gamma^{\prime})/5, (18-\gamma^{\prime})/5, \gamma^{\prime}]$ model 
(equation~\ref{EqDMc3})
from Dehnen \& McLaughlin (2005, their equation~46b) 
to our dark matter density profiles.  
}
\label{FigDMc3}
\end{figure}

As shown in Figure 10, for three of the six cluster-sized halos, 
this model has the greatest
residual scatter of the four, 3-parameter models tested here.  For
another two of the six cluster-sized halos it has the second greatest
residual scatter.  This model is also unable to match the curvature in
the halos of the cold collapse models (Figure~\ref{FigDMc3}).  However,
it does provide very good fits to the galaxy-sized halos, and actually has
the smallest residual scatter for three of these halos 
(Table~\ref{Table1b}).

The shallowest, inner, negative logarithmic slope of this model occurs
when $\beta_0 = 0$, giving a value of $7/9 \approx 0.78$.  For
non-zero values of $\beta_0$, this slope steepens roughly linearly
with $\beta_0$.

\subsection{S\'ersic/Einasto model} \label{SecSer}

S\'ersic (1963, 1968) generalized de Vaucouleurs' (1948) $R^{1/4}$ 
luminosity profile model 
by replacing the exponent $1/4$ with $1/n$, such that $n$ was a free 
parameter that measured the `shape' of a galaxy's luminosity profile.
Using the observers' notion of `concentration' 
(see the review in Graham, Trujillo, \& Caon 2001), 
the quantity $n$ is monotonically related to how centrally concentrated 
a galaxy's light profile is. 
With $R$ denoting the {\it projected} radius, 
S\'ersic's $R^{1/n}$ model is often written as 
\begin{equation}
I(R)=I_{\rm e} \exp\left\{ -b_n\left[ (R/R_{\rm e}) ^{1/n} -1\right] \right\},
\label{Sersic}
\end{equation}
where $I_{\rm e}$ is the (projected) intensity at the (projected)
effective radius $R_{\rm e}$.  The term $b_n$ is not a parameter but a
function of $n$ and defined in such a way that $R_{\rm e}$ encloses
half of the (projected) total galaxy light (Caon et al.\ 1993; see
also Ciotti 1991, his equation (1)).  A good approximation when $n
\gtrsim 0.5$ is given in Prugniel \& Simien (1997) as 
\begin{equation}
b_n \approx 2n-1/3+0.009876/n.
\label{Eqbutt}
\end{equation}
Assorted expressions related to the $R^{1/n}$ model can be found in
Graham \& Driver's (2005) review article.

Despite the success of this model in describing the luminosity profiles
of elliptical galaxies (e.g., Phillipps et al.\ 1998; 
Caon et al.\ 1993; D'Onofrio et al.\ 1994; Young \& Currie 1995; 
Graham et al.\ 1996; Graham \& Guzm\'an 2003, and references therein), 
it is nonetheless an empirical fitting function with no 
commonly recognized physical basis.
We are therefore free to explore the suitability of this function for
describing the {\it mass} density profiles, $\rho(r)$, of dark matter halos.
Indeed, Einasto (1965, eq.~4; 1968, eq.~1.7; 1969, eq.~3.1) 
independently developed  the functional form of S\'ersic's equation 
and used it to describe density profiles. 
More recent application of this profile to the modelling of
density profiles can be found in
Einasto \& Haud (1989, their eq.~14) and 
Tenjes, Haud, \& Einasto (1994, their eq.~A1).
Most recently, the same model has been applied by 
Navarro et al. (2004) and Merritt et al. (2005)
to characterize dark matter halos, and 
Aceves, Vel\'azquez, \& Cruz's (2006) used it to
describe merger remants
in simulated disk galaxy collisions. 

To avoid potential confusion with S\'ersic's $R^{1/n}$ model, 
we define the following expression as ``Einasto's $r^{1/n}$ model'':
\begin{equation}
\rho(r)=\rho_{\rm e} \exp\left\{ -d_n\left[ (r/r_{\rm e})^{1/n} -1\right] \right\},
\label{SerDen}
\end{equation}
where $r$ is the {\it spatial} (i.e., not projected) radius.  The term
$d_n$, defined below, is a function of $n$ such that $\rho_{\rm e}$ is
the density at the radius $r_{\rm e}$ which defines a volume
containing half of the total mass.  
The central density is finite and given by 
$\rho(r=0) = \rho_{\rm e} {\rm e}^{d_n}$.

The integral of equation~(\ref{SerDen}) over some volume gives the
enclosed mass\footnote{A similar expression is given in Mamon \&
{\L}okas 2005, their equation~(A2); and Cardone et al.\ 2005, their
equation~(11).}, which is also finite and equal to
\begin{equation}
M(r) = 4\pi \int_0^r \rho({\bar r}) {\bar r}^2 {\rm d}{\bar r}. 
\end{equation}
This can be solved by using the substitution ${\bar x} \equiv d_n({\bar r}/r_{\rm e})^{1/n}$
to give
\begin{equation}
M(r) = 4\pi n r_{\rm e}^{3} \rho_{\rm e} {\rm e}^{d_n} {d_n}^{-3n} \gamma (3n,x),
\label{Sermass}
\end{equation}
where $\gamma(3n,x)$ is the incomplete gamma function defined by
\begin{equation}
\gamma (3n,x)=\int ^{x}_{0} {\rm e}^{-t}t^{3n-1} {\rm d}t.
\label{gamFunc}
\end{equation}
Replacing $\gamma (3n,x)$ with $\Gamma (3n)$ in equation~(\ref{Sermass})
gives the total mass $M_{tot}$. 

\begin{figure}
\includegraphics[scale=0.5,angle=270]{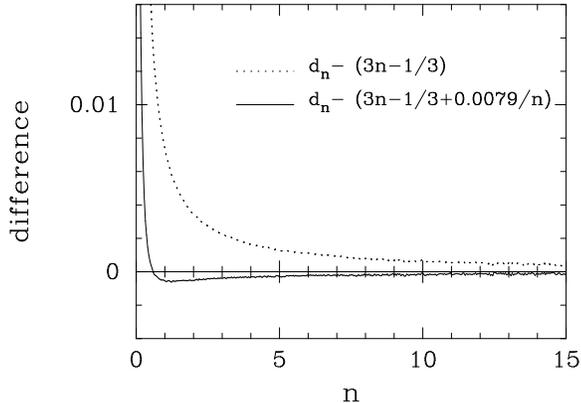}
\caption{
Difference between the exact value for $d_n$ from equation~(\ref{SerDen}), 
such that $\Gamma (3n)=2\gamma (3n,d_n)$, and the two approximations inset 
in the Figure.
}
\label{Figd}
\end{figure}

The value of $d_n$, which we first saw in equation~(\ref{SerDen}), is
obtained by solving $\Gamma (3n)=2\times\gamma (3n,d_n)$, where $\Gamma $ is
the (complete) gamma function.  
The value of $d_n$ can be well approximated (Mamon 2005, priv.\ comm.) 
by the expression 
\begin{equation}
d_n \approx 3n - 1/3 + 0.0079/n, \hbox{ for } n \gtrsim 0.5
\label{Eqdud}
\end{equation} 
(see Figure~\ref{Figd}). 

In Paper II we recast Einasto's $r^{1/n}$ model using the radius $r_{-2}$,  
where the logarithmic slope of the density profile equals $-2$. 

Einasto's $r^{1/n}$ model (see Einasto \& Haud 1989)
was used in Navarro et al.\ (2004; their 
equation~5) to fit their simulated dark matter halos. 
They obtained $n \approx 1/(0.172\pm0.032) \approx 6\pm1.1$.  
Subsequently, Merritt et al.\ (2005) showed that Einasto's $r^{1/n}$  
model performed as well as the (1, 3, $\gamma$) model, and gave better fits
for the dwarf- and galaxy-sized halos, obtaining $n\approx 5.6\pm0.7$.
For a sample of galaxy-sized halos, Prada et al.\ (2005) obtained
similar values of $6-7.5$.

\begin{figure}
\includegraphics[scale=0.40,angle=0]{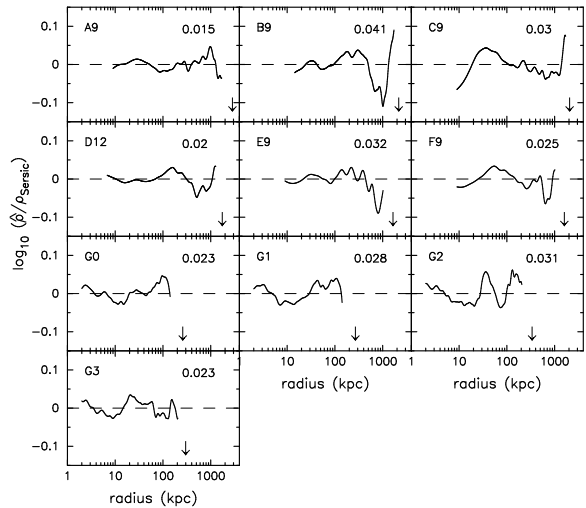}
\caption{ 
 Residual profiles from 
application of Einasto's $r^{1/n}$ model (equation~\ref{SerDen}) to our dark
matter density profiles.
}
\label{FigSer}
\end{figure}

Figure~\ref{FigSer} shows the application of equation~(\ref{SerDen}) to
the $N$-body halos of Section~\ref{Dave}.  
A comparison
with the (1, 3, $\gamma$) model fits in Figure~\ref{FigNFW} 
reveals that Einasto's model
provides a better description for 
five of the six cluster-sized halos, 
three of the four galaxy-sized halos, 
and both of the spherical collapse halos.  

Navarro et al.\ (2004) wrote ``adjusting the parameter [$n$] allows
the profile to be tailored to each individual halo, resulting in
improved fits''\footnote{The value of $n$, equal to $1/\alpha$ in
Navarro et al.'s (2004) notation, ranged from 4.6 to 8.2 (Navarro et
al.\ 2004, their table~3).}.  Such a breaking of structural homology
(see Graham \& Colless 1997 for an analogy with projected
luminosity profiles) replaces the notion that a universal density profile
may exist.

A number of useful expressions pertaining to Einasto's model, when
used as a density profile (equation~\ref{SerDen}), are given in
Cardone et al.\ (2005) and Mamon \& {\L}okas (2005).  In particular,
Cardone et al.\ provide the gravitational potential, as well as
approximations to the surface density and space velocity dispersion of
the Einasto $r^{1/n}$ model, while Mamon \& Lokas give approximations
for the concentration parameter, central density, and $M_{\rm
virial}/M_{\rm total}$.  The nature of the inner profile slope of
Einasto's $r^{1/n}$ model and several other useful quantities are
presented in Paper II. 

\subsection{Prugniel-Simien model: 
A deprojected S\'ersic $R^{1/n}$ model}\label{SecPS}

Merritt et al.\ (2005) tested how well a
deprojected S\'ersic $R^{1/n}$ model fit $\rho(r)$ from
the Navarro et al. (2004) $N$-body halos.
This was essentially the same as comparing the 
halo {\it surface} densities with S\'ersic's $R^{1/n}$ law.
Prugniel \& Simien (1997) presented
a simple, analytical approximation to the deprojected
S\'ersic law (their eq. B6):
\begin{equation}
  \rho(r) = \rho^{\prime} \left({r\over R_{\rm e}}\right)^{-p}
             \exp\left[-b\left( r/R_{\rm e} \right)^{1/n} \right], 
\label{EqPS97}
\end{equation}
with
\begin{equation} \label{projIe}
\rho^{\prime} = {M\over L} \ I_{\rm e} {\rm e}^b \ b^{n(1-p)} \ 
{\Gamma(2n) \over 2 R_{\rm e} \Gamma(n(3-p)) }.
\label{EqPS97b}
\end{equation}
Equation~(\ref{EqPS97}) is a generalization of equation~(2) in 
Mellier \& Mathez (1987), who considered only approximations
to the deprojected $R^{1/4}$ law.
Mellier \& Mathez's model was itself a 
modification of equation~(33) from Young (1976), 
which derived from the work of Poveda, Iturriaga, \& Orozco (1960).

In these expressions, $R_e$, $n$ and $b$ are understood to be
essentially the same quantities that appear in the S\'ersic $R^{1/n}$
law that describes the projected density (equation~\ref{Sersic}).
In fact, since equation~(\ref{EqPS97}) is not exactly
a deprojected S\'ersic profile, the correspondence between the
parameters will not be perfect.
We follow the practice of earlier authors and define $b$ 
to have the same relation to $n$ as in equation~(\ref{Eqbutt}).
(For clarity, we have dropped the subscript $n$ from $b_n$.)
Although the parameter
$\rho^{\prime}$ is obtained from fitting the density profile, it can be 
defined in such a way that the total (finite) mass
from equation~(\ref{EqPS97}) equals that from equation~(\ref{Sersic}), 
giving equation~(\ref{projIe}). 
(We stress that the $n$ in the Prugniel-Simien profile
is {\it not} equivalent to the $n$ 
in equation~(\ref{SerDen}), Einasto's model.)

\begin{figure}
\includegraphics[scale=0.48]{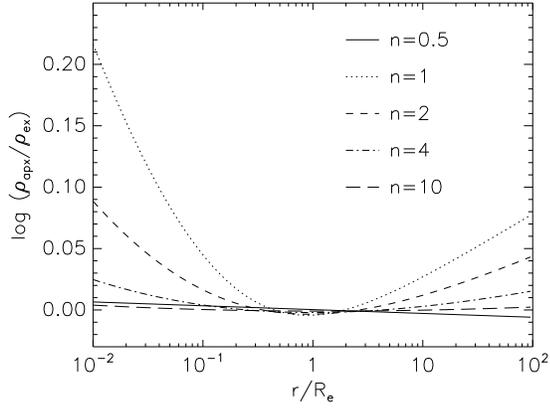}
\caption{
Logarithmic difference between the exact deprojection of
S\'ersic's $R^{1/n}$ model (equation~\ref{Sersic}) and the approximation
given by Prugniel \& Simien (1997) in equation~(\ref{EqPS97}), using
the values of $p$ and $b$ given in equations~(\ref{pvalue}) and 
(\ref{Eqbutt}),
respectively. 
}
\label{Balsa}
\end{figure}

This leaves the parameter $p$.
We define $p$,
like $b$, uniquely in terms of $n$:
\begin{equation}
p = 1.0 - 0.6097/n + 0.05463/n^2 .
\label{pvalue}
\end{equation} 
Lima Neto et al.\ (1999) derived this expression by requiring
the projection of equation~(\ref{EqPS97}) to approximate
as closely as possible to the S\'ersic profile with the same
$(R_e,n)$,
for $0.6\le n\le 10$ and $10^{-2}\le R/R_e\le 10^3$.\footnote{The 
value of $p$ given in equation~(\ref{pvalue}) is
preferable to the value $1.0 - 0.6097/n + 0.05563/n^2$ given in 
M\'arquez et al.\ (2000), (Lima Neto 
2005, priv.comm.).}
The accuracy of Prugniel \& Simien's (1997) approximation,
using equation~(\ref{pvalue}) for $p(n)$, is shown in 
Figure~\ref{Balsa}.

Terzi\'c \& Graham (2005) give
simple expressions, in terms of elementary and special
functions, for the gravitational potential and force
of a galaxy obeying the Prugniel-Simien law, and
derive the spatial and line-of-sight velocity dispersion 
profiles.

One could also allow $p$ to be a free parameter,
creating a density profile that has any desired inner slope.
For instance, setting $p=0$, the Prugniel-Simien model reduces
to the Einasto model.
We do not explore that idea further here.

The density at $r=R_{\rm e}$ is given by $\rho_{\rm e}=\rho^{\prime} {\rm
e}^{-b}$, while the projected surface density at $R=R_{\rm e}$,
denoted by $I_{\rm e}$, can be solved for using equation~(\ref{projIe}).
Thus, one can immediately construct (a good approximation to) the
projected mass distribution, which will have a S\'ersic form
(equation~\ref{Sersic}) with parameters ($R_{\rm e}, I_{\rm e}$, and
$n$).
This allows the halo parameters to be directly compared
with those of S\'ersic fits to luminous galaxies,
which we do in Paper III.
In Paper II we recast this model using the radius 
where the logarithmic slope of the density profile equals $-2$. 

The mass profile (Terzi\'c \& Graham 2005, their Appendix A; see also
Lima Neto et al.\ 1999 and M\'arquez et al.\ 2001), can be written as
\begin{equation} \label{PSmass}
M(r)  = {4\pi \rho^{\prime} {R_{\rm e}}^3n} {b^{n(p-3)}}
\gamma\left(n(3-p),Z\right),
\end{equation}
where $Z \equiv b(r/R_{\rm e})^{1/n}$ and $\gamma(a,x)$ is the incomplete gamma
function given in equation~(\ref{gamFunc}).
The total mass is obtained by replacing $\gamma\left(n(3-p),Z\right)$
with $\Gamma\left(n(3-p)\right)$, and the circular velocity is 
given by $v_{\rm circ}(r) = \sqrt{ GM(r)/r }$. 

\begin{figure}
\includegraphics[scale=0.40,angle=0]{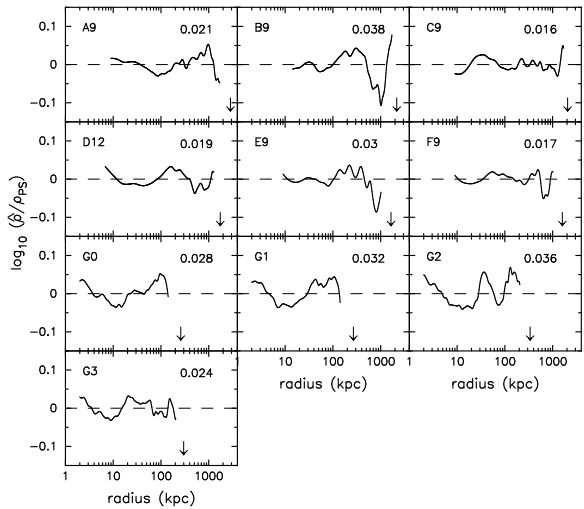}
\caption{
 Residual profiles from 
application of the Prugniel \& Simien model (equation~\ref{EqPS97})
to our dark matter density profiles.  
%
}
\label{FigPS97}
\end{figure}

In Figure~\ref{FigPS97}, equation~(\ref{EqPS97}) has been applied to our
dark matter profiles. 
The average ($\pm$ standard deviation) of the shape parameter for the
galaxy-sized and cluster-sized halos is $n=3.59 (\pm 0.65)$ and
$n=2.89 (\pm 0.49)$, respectively.
Merritt et al.\ (2005, their Table~1) found values of 3.40$\pm$0.36
and 2.99$\pm$0.49 for their sample of galaxies and clusters, 
respectively, in good agreement with the results obtained here using a
different set of $N$-body simulations and equation~(\ref{EqPS97}),
rather than a numerically deprojected $R^{1/n}$ light profile.

Figure~\ref{FigPS97} reveals that CDM halos 
resemble galaxies (Merritt et al. 2005), since
the projection of the Prugniel-Simien model closely
matches the S\'ersic $R^{1/n}$ model, and the latter
is a good approximation to the luminosity profiles
of stellar spheroids.
Subject to vertical and horizontal scaling, 
CDM halos have similar mass distributions to
elliptical galaxies with an absolute $B$-band 
magnitude around $-18 \pm
 1$ mag; 
these galaxies have $n\sim 3$ 
(see Graham \& Guzm\'an 2003, their figure~9).  
This result was obscured until recently due to the use
of different empirical models by observers and modelers.

Before moving on, we again remark that we have not explored potential
refinements to the expression (\ref{pvalue}) for the quantity $p$, 
but note that this could result in a better matching of the model to 
the simulated profiles at small radii.
As the resolution of $N$-body clustering simulations continues
to improve,
it will make sense to explore such generalizations.

\section{Model comparison: Which did best?}

Table~\ref{Table3} summarizes how well each parametric
model performed by listing, for each type of halo, 
the rms value of $\Delta$ 
(equation~\ref{EqChi}) for each set of halos,
given by 
\begin{equation}
\Delta_{\rm rms} = \sqrt{ \frac{1}{N} \sum_{i=1}^{N} \Delta_i^2 }, 
\label{EqDelrms}
\end{equation}
with $N$ = 6, 4 and 2 for the cluster-sized, galaxy-sized, and 
spherical-collapse halos, respectively. 
%
%
A detailed description of each model's performance follows. 

The bowl- and hump-shaped residual profiles associated with the
2-parameter model of Burkert (1995) reveal this model's 
inability to describe the radial mass distribution in our 
simulated dark matter halos.
The 2-parameter model of Dehnen \& McLaughlin (2005)
performs considerably better, although it too fails to describe the
cold collapse systems and two of the six cluster-sized halos, 
specifically C09 and F09.  Although this (4/9, 31/9, 7/9) 
model never provides the best 
fit, it does equal or out-perform the NFW-like (1, 3, $\gamma$) model
in describing 3 of the 12 halos (A09, D12, G03).  

In general, all of the 3-parameter models perform well ($0.015
\lesssim \Delta \lesssim 0.04$ dex) at fitting the $N$-body
(non-collapse) halos.
However, neither the (1, 3, $\gamma$) model nor the
3-parameter Dehnen-McLaughlin model 
can match the curvature in the
density profiles of the cold collapse systems (M11 \& M35).  On the
other hand, both Einasto's $r^{1/n}$ model and that from Prugniel \& Simien
give reasonably good fits ($\Delta \sim 0.05$ dex) for these two
halos.

\begin{figure}
\includegraphics[scale=0.4,angle=270]{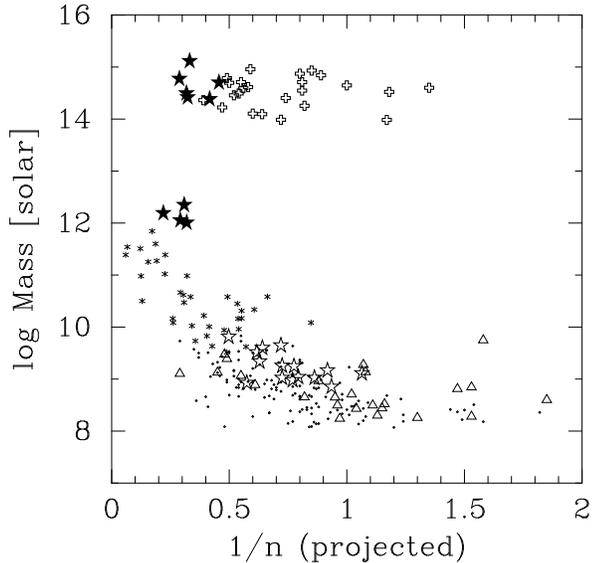}
\caption{
Mass versus profile shape ($1/n$). 
For the galaxies and galaxy clusters, the shape parameters $n$ have come from
the best-fitting S\'ersic $R^{1/n}$ model to the (projected) luminosity-
and X-ray profiles, respectively. 
The galaxy stellar masses, and cluster gas masses are shown here.  
For DM halos, the virial masses are shwon and the 
shape parameters have come from  the best-fitting
Prugniel-Simien model.
(Note: The value of $1/n$ from the Prugniel-Simien model
applied to a density profile is equivalent to the value of $n$ from 
S\'ersic's model applied to the projected distribution.) 
We are plotting baryonic properties for the galaxies
alongside dark matter properties for the simulated halos.
Filled stars: $N$-body, dark matter halos from this paper;
open plus signs: galaxy clusters from Demarco et al.\ (2003);
dots: dwarf Elliptical (dE) galaxies from Binggeli \& Jerjen (1998);
triangles: dE galaxies from Stiavelli et al.\ (2001);
open stars: dE galaxies from Graham \& Guzm\'an (2003); 
asterisk: intermediate to bright elliptical galaxies from Caon et al.\ 
(1993) and D'Onofrio et al.\ (1994). 
}
\label{FigM_n}
\end{figure}

The Prugniel-Simien model provided the best overall description of the
cluster-sized, $N$-body halos.  The (1, 3, $\gamma$) model and the
3-parameter Dehnen-McLaughlin model provided the best fit for only one
cluster-sized, $N$-body halo each, and even then the (1, 3, $\gamma$)
model only just out-performed the Prugniel-Simien model which gave the
best fit for four of the six cluster-sized halos. For two of these
halos, the size of the residual about the optimal Prugniel-Simien fit
was roughly half of the value obtained when using the (1, 3, $\gamma$)
model.

The implication of this result is that S\'ersic's $R^{1/n}$ model will
describe the projected surface density of the cluster-sized, dark
matter halos.  Intriguingly, Demarco et al.\ (2003) and Durret, 
Lima Neto \& Forman (2005) have observed that
the (projected) hot X-ray gas distribution in clusters can indeed be 
described with S\'ersic's $R^{1/n}$ model; although the gas can at times
display a rather unrelaxed behavior (Statler \& Diehl 2006).  
Studies of gravitational
lensing may therefore benefit from the use of S\'ersic's $R^{1/n}$
model for which the lensing equation has been solved (Cardone et al.\
2004) and for which numerous other properties have previously been
computed (Graham \& Driver 2005).

With regard to the galaxy-sized, $N$-body halos, the situation is 
somewhat different. 
Dehnen \& McLaughlin's (2005) anisotropic 3-parameter model provided
the best fit for three of the four profiles, with 
the Einasto $r^{1/n}$ model providing the best fit for the forth profile. 
We also observe that Einasto's model 
provided better fits than the (1, 3, $\gamma$) model for three of 
the four $N$-body halos. 
If this observation holds, namely, that the Prugniel-Simien model
describes the density profiles of the cluster-sized halos best, while
Dehnen \& McLaughlin's 3-parameter model 
provides the best description of the galaxy-sized 
halos, it would imply that these halos do not have the same structural
form.  
Of course, even if the same model {\it did} provide the best fit
for both types of halo, any variation in the value of the profile
shape $n$, or central isotropy parameter $\beta_0$, 
would point toward the existence of nonhomology.

\begin{figure}
\includegraphics[scale=0.43,angle=270]{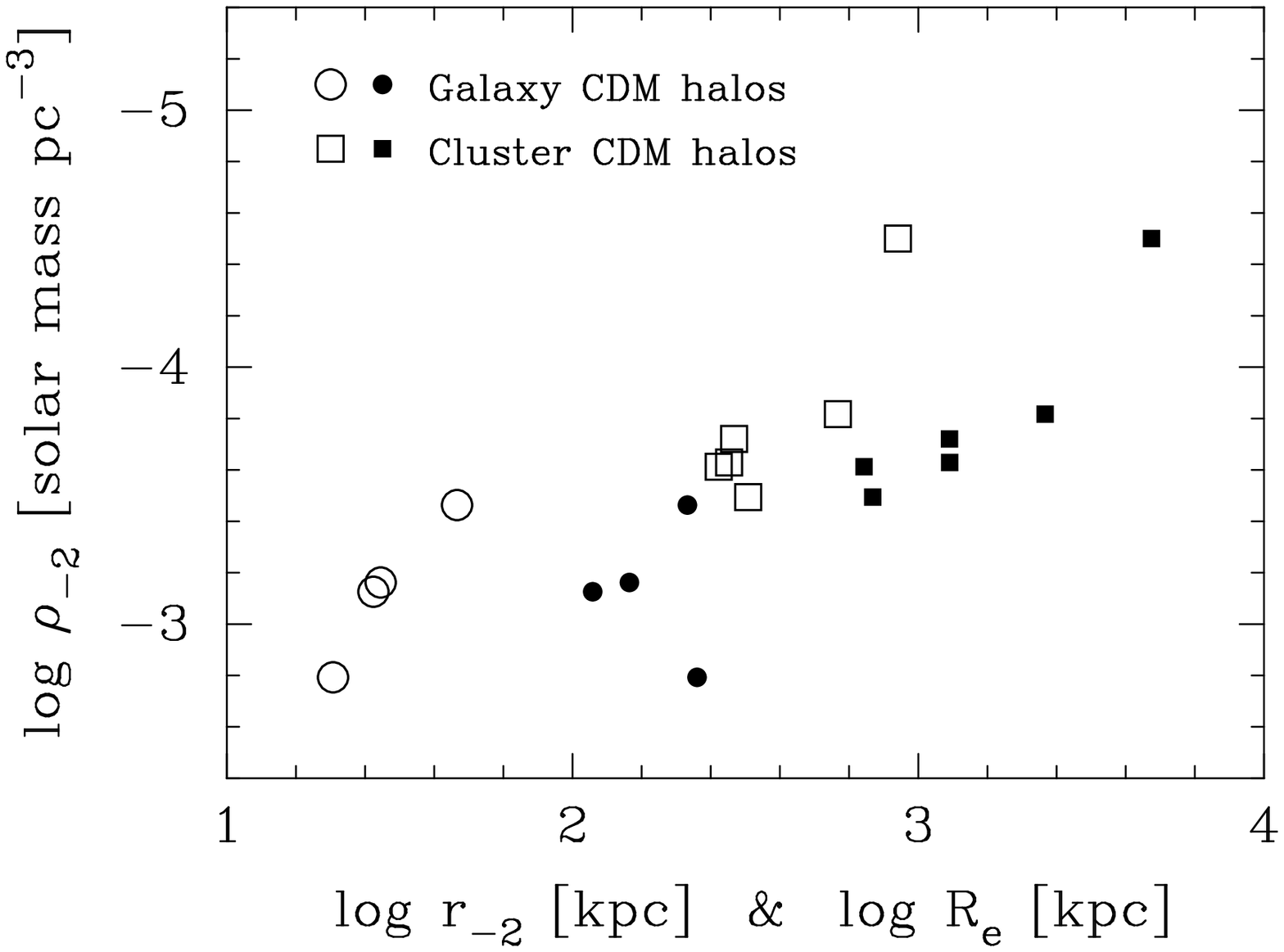}
\caption{
The density, $\rho_{-2}$, where the logarithmic slope of the density
profile equals $-2$ is plotted against i) the radius where this occurs
(open symbols), and ii) the effective radius (filled symbols) derived
from the best-fitting Prugniel-Simien model (equation~\ref{EqPS97}).
Both $\rho_{-2}$ and $r_{-2}$ are also computed from the best-fitting
Prugniel-Simien model, see Paper II.  If a universal profile existed for
these halos, then the vertical difference should be constant for all halos. 
}
\label{PS_R_2}
\end{figure}

While halos of different mass may be systematically
better described by different density laws, 
it is important to emphasize that a {\it single} density law provides
a good fit to {\it all} of the $N$-body halos considered here.
As Table~4 shows, Einasto's $r^{1/n}$ law has the smallest,
or second-smallest, value of $\Delta_{rms}$ for galaxy-sized,
cluster-sized, and spherical-collapse halos.
None of the other parametric models which we considered
performs as well ``across the board.''
The next best performer overall is the Prugniel-Simien profile.

\section{Discussion}

Figure~\ref{FigM_n} shows our $N$-body halos, together with 
real elliptical galaxies and clusters, in the profile shape vs. mass plane. 
The profile shape parameter plotted there is either $n$ from the
S\'ersic $R^{1/n}$ model fit to the light profile,
or the corresponding parameter from the Prugniel-Simien model fit
to the dark-matter density.
Dynamical masses from the Demarco et al.\ (2003) study of 
galaxy clusters are shown.
We have also included the elliptical 
galaxy compilation in Graham \& Guzm\'an (2003), converting their
$B$-band luminosities into solar masses using 
a stellar mass-to-light ratio of 5.3 (Worthey 1994, for a 12 Gyr old SSP),
and an absolute $B$-band magnitude for the Sun of 5.47 $B$-mag (Cox 2000). 
This approach ignores the contribution from dark matter in galaxies. 
However, given the uncertainties on how $M_{\rm tot}/L$ varies with $L$
(e.g., Trujillo, Burkert, \& Bell 2004, and references therein) we 
prefer not to apply this correction, and note that the galaxy masses 
in Figure~\ref{FigM_n} only reflect the stellar mass. 

Figure~\ref{FigM_n} suggests that the simulated galaxy-sized
halos have a different shape parameter, i.e. a different mass 
distribution, than the simulated cluster-sized halos.
The same conclusion was reached by Merritt et al.\ (2005)
who studied a different sample of $N$-body halos.
The sample of dwarf- and galaxy-sized halos from that paper
had a mean ($\pm$ standard deviation)\footnote{Reminder: the 
uncertainty on the mean is not equal to the standard deviation.} 
profile shape $n=3.04 (\pm 0.34)$, 
while the cluster-sized halos had $n=2.38 (\pm 0.25)$. 
We observe this same systematic difference in our $N$-body halos. 
Taking the profile shape $n$ from the Prugniel-Simien model fits to
the density profile (equivalent to the value of $n$ obtained by 
fitting S\'ersic's $R^{1/n}$ model to the projected distribution) 
we find $n=3.59 (\pm0.65)$ for our cluster-sized halos and 
$n=2.89 (\pm0.49)$ for our galaxy-sized halos.
A Student $t$ test, without assuming equal
variance in the two distributions, reveals the above means are
different at the 88\% level.  Applying the same test to the data of
Merritt et al.\ (2005; their Table~1, column 2), which is double the
size of our sample and also contains dwarf galaxy-sized halos, we find
that the means are different at the 99.98\% level.  
We conclude that there is a significant mass dependence in the
density profiles of simulated dark-matter halos.
Density profiles of more massive halos exhibit more curvature
(smaller $n$) on a log-log plot.

The fact that $n$ varies systematically with halo mass raises 
the question of which density scale and radial scale to use
when characterizing halo structure.
In the presence of a ``universal'' density profile, the ratio
between $R_{\rm e}$ and $r_{-2}$ (the radius where the
logarithmic slope of the density profile equals $-2$, see
Paper II) is a constant factor, but with varying values of
$n$ this is not the case.  This remark also holds for the scale
density, which is used to measure the contrast with the background
density of the universe and provides the so-called ``halo
concentration.''
This in turn raises the question of what ``concentration'' should 
actually be used, and whether systematic biases exist if one uses 
$\rho_{-2}$ 
 rather than, say, $\rho_{\rm e}$.   
To reiterate this point: the density ratio between $r=r_{-2}$ 
and $r=R_{\rm e}$ depends on the profile
shape $n$, and thus, apparently, on the halo mass.  

In Figure~\ref{PS_R_2} we show how the use of $r_{-2}$ and $R_{\rm e}$
produce slightly different results in the size-density diagram (e.g.,
Figure 8 of Navarro et al. 2004). 
The relation between size (or equivalently mass) and central
concentration (or density) varies depending on how one
chooses to measure the sizes of the halos.

To better explore how the homology (i.e., universality) of 
CDM halos is broken,
it would be beneficial to analyze a large, low-resolution sample of
halos from a cosmological cube simulation in order to obtain good
statistics. 
Moreover, the
collective impact from differing degrees of virialization in the outer
regions, possible debris wakes from larger structures, global ringing
induced by the last major merger, triaxiality, and the presence of
large subhalos could be quantified.

\begin{deluxetable}{lccc}
\tablewidth{0pt}
\tabletypesize{\footnotesize}
\tablecaption{Residual scatter: rms values of $\Delta$.}
\tablehead{ 
\colhead{Model}  &  \colhead{Cluster-sized} & \colhead{Galaxy-sized} & \colhead{Spherical-collapse}  \\
\colhead{ } &  \colhead{halos} & \colhead{halos}  & \colhead{halos} 
}
\startdata
 \multicolumn{4}{c}{3-parameter models} \\
 Einasto                            & {\bf 0.028} &  {\bf 0.026}  & {\bf 0.052} \\
Prugniel-Simien                     & {\bf 0.025} &  0.030        & {\bf 0.056} \\
 $(1,3,\gamma)$                     & 0.032       &  0.028        & 0.236 \\
Dehnen-McLaughlin (Eq.\ref{EqDMc3}) & 0.034       &  {\bf 0.023}  & 0.193 \\
 \multicolumn{4}{c}{2-parameter models} \\
Dehnen-McLaughlin (Eq.\ref{EqDMc2}) & 0.053       & 0.032         & 0.257 \\
NFW                                 & 0.046       & 0.046         & 0.246 \\ 
Burkert                             & 0.218       & 0.164         & 0.217 \\
\enddata
\tablecomments{
Col.(1): Model.  
Col.(2): rms of the 6 residual scatters, $\Delta_{\rm rms}$ (equation~\ref{EqDelrms}), 
for the cluster-sized halos. 
Col.(3): Similar to Col.(2) but for the 4 galaxy-sized halos.
Col.(3): Similar to Col.(2) but for the 2 spherical collapse halos. 
For each halo type, the two models which perform the best are highlighted in bold. 
\label{Table3}
}
\end{deluxetable}


\section{Summary} \label{SecSum}

We presented a nonparametric algorithm for extracting smooth
and continuous representations of spherical density profiles
from $N$-body data, and applied it to a sample of simulated,
dark matter halos.
All halos exhibit a continuous variation of logarithmic
density slope with radius; in the case of the $\Lambda$CDM
halos, the variation of slope with radius is close to a power
law.
We then compared the ability of a variety of parametric
models to reproduce the nonparametric $\rho(r)$'s.
Over the fitted radial range $0.01 \lesssim r/r_{\rm vir} <1$, 
both the Einasto $r^{1/n}$ model (identical in functional
form to S\'ersic's model but expressed in terms of space, 
rather than projected, radius and density)
and the Prugniel-Simien model (an analytical approximation 
to a de-projected S\'ersic law)
provide a better description of the data than the
(1, 3, $\gamma$) model, 
i.e.\ the NFW-like double power-law model
with inner slope $\gamma$. 
Moreover, unlike the (1, 3, $\gamma$) model,
both of these models have finite total mass,
and are also capable of
describing the density profiles of halos formed from the cold collapse
of a spherical over-density (Figure~\ref{Figcold}). 

The single function that provides the best overall fit
to the halo density profiles is Einasto's law, equation~(\ref{SerDen}):
$$
\rho(r)=\rho_{\rm e} \exp\left\{ -d_n\left[ (r/r_{\rm e})^{1/n} -1\right] 
\right\} \nonumber
$$
with $d_n$ defined as in equation~(\ref{Eqdud}).
This conclusion is consistent with that of an earlier
study (Merritt et al. 2005) that was based on a different
set of $N$-body halos.
Typical values of the ``shape'' parameter $n$ 
in equation~(\ref{SerDen}) are
$4\lesssim n\lesssim 7$ (Table 1).
Corresponding $n$ values from S\'ersic profile fits
to the projected (surface) density range from
$\sim 3$ to $\sim 3.5$ (Fig.~\ref{FigM_n}).

We propose that Einasto's model, equation~(\ref{SerDen}),
be more widely used to characterize the density
profiles of $N$-body halos.
As noted above, Einasto's model has already found application in a number
of observationally-motivated studies of the distribution of
mass in galaxies and galaxy clusters.
We propose also that the suitability of Einasto's model 
for describing the luminous density profiles of galaxies
should be evaluated -- either by projecting equation~(\ref{SerDen})
onto the plane of the sky, or by comparing equation~(\ref{SerDen})
directly with deprojected luminosity profiles.
Such a study could strengthen the already strong connection
between the density profiles of galaxies and $N$-body
dark-matter halos (Merritt et al. 2005).

While equation~(\ref{SerDen}) is a good description of
all of the halo models considered here, we found that
systematic differences do exist in the best-fit models that describe
$N$-body halos formed via hierarchical merging on the one hand,
and those formed via spherical collapse on the other hand, 
in the sense that the latter have substantially smaller shape 
parameters, $n\approx 3.3$ (Table 1).
That is, the density profiles in the cold collapse
halos decline more quickly than $r^{-3}$ at large radii, and have
shallower inner profile slopes than those produced in simulations of
hierarchical merging.  

With regard just to the non-collapse models, we also found
systematic differences between the cluster- and galaxy-sized halos.
The latter are slightly better fit by the 3-parameter
Dehnen-McLaughlin model, 
and the former are slightly better fit by the Prugniel-Simien
model (Table 4).
This, together with the observation that more massive halos tend
to have smaller shape parameters $n$ (Figure~\ref{FigM_n}), 
suggests that there may not be a truly ``universal'' density
profile that describes $\Lambda$CDM halos.

\acknowledgments  

We kindly thank Gary Mamon for his detailed comments on this
manuscript, as well as a second, anonymous referee.  
We are additionally grateful to Walter Dehnen and Dean
McLaughlin for their helpful corrections and comments, and to Carlo
Nipoti and Luca Ciotti.  We also wish to thank Peeter Tenjes for
tracking down and kindly faxing us copies of Einasto's original papers
in Russian.
A.G.\ acknowledges support from NASA grant HST-AR-09927.01-A
from the Space Telescope Science Institute, 
and the Australian Research Council through Discovery Project Grant DP0451426.
D.M.\ was supported by grants AST 04-20920 and AST
04-37519 from the National Science Foundation and grant NNG04GJ48G
from NASA.
%
%
J.D.\ is grateful for financial support from the Swiss National
Science Foundation.
B.T.\ acknowledges support from Department of Energy grant G1A62056.


\begin{references}
\reference{Aar63}Aarseth, S.J.\ 1963, MNRAS, 126, 223
\reference{Abdelsalam98} Abdelsalam, H.~M., 
Saha, P., \& Williams, L.~L.~R.\ 1998, \aj, 116, 1541 
\reference{AVC06}Aceves, H., Vel\'azquez, H., \& Cruz, F.\ 2006, MNRAS, submitted (astro-ph/0601412)
\reference{AaM90}Aguilar, L.A., \& Merritt, D.\ 1990, ApJ, 354, 33
\reference{Aet05}Austin, C.G., Williams, L.L.R., Barnes, E.I., Babul, A., \& Dalcanton, J.J.\ 2005, ApJ, 634, 756
\reference{Ber85}Bertschinger, E.\ 1985, ApJS, 58, 39
\reference{BaJ98}Binggeli, B., \& Jerjen, H.\ 1998, A\&A, 333, 17
\reference{Binney82}Binney, J. J. 1982,
  MNRAS, 200, 951
\reference{Bur95}Burkert, A.\ 1995, ApJ, 447, L25 
\reference{CCD93}Caon, N., Capaccioli, M., \& D'Onofrio, M.\ 1993, MNRAS, 265, 1013
\reference{Car04}Cardone, V.F.\ 2004, A\&A, 415, 839
\reference{CPT05}Cardone, V.F., Piedipalumbo, E., \& Tortora, C.\ 2005, MNRAS, 358, 1325
\reference{Cio91}Ciotti, L.\ 1991, A\&A, 249, 99
\reference{Cox00} Cox, D.P.\ 2000, Allen's Astrophysical quantities, New York: AIP Press; Springer
\reference{CH74} Cox, D. R. \& Hinkley, D. V. 1974,
Theoretical Statistics, London: Chapman and Hall
\reference{DaH01}Dalcanton, J.D., \& Hogan, C.J.\ 2001, ApJ, 561, 35
\reference{DMc05}Dehnen, W., \& McLaughlin, D.E.\ 2005, MNRAS, 363, 1057
\reference{Dem03}Demarco, R., Magnard, F., Durret, F., \& M\'arquez, I.\ 2003, A\&A, 407, 437
\reference{deV48}de Vaucouleurs, G.\ 1948, Ann.\ d'astrophys., 11, 247
\reference{DMS4a}Diemand, J., Moore, B., \& Stadel, J.\ 2004a, MNRAS, 352, 535 
\reference{DMS4b}Diemand, J., Moore, B., \& Stadel, J.\ 2004b, MNRAS, 353, 624 
\reference{DZMSC05} Diemand, J., Zemp, M., 
Moore, B., Stadel, J., \& Carollo, C.~M.\ 2005, \mnras, 364, 665 
\reference{DCC94}D'Onofrio, M., Capaccioli, M., \& Caon, N.\ 1994, MNRAS, 271, 523
\reference{DaC91}Dubinski, J., \& Carlberg, R.\ 1991, ApJ, 378, 496
\reference{DLNF5}Durret, F., Lima Neto, G.H., \& Forman, W.\ 2005, A\&A, 432, 809
\reference{Eet88}Efstathiou, G.P., Frenk, C.S., White, S.D.M., \& Davis, M.\ 1988, MNRAS, 235, 715
\reference{Ein65}Einasto, J.\ 1965, Trudy Inst.\ Astrofiz.\ Alma-Ata, 5, 87
\reference{Ein68}Einasto, J.\ 1968, Tartu Astr.\ Obs.\ Publ.\ Vol.\ 36, Nr 5-6, 414,
\reference{Ein69}Einasto, J.\ 1969, Astrofizika, 5, 137
\reference{EaH89}Einasto, J., \& Haud, U.\ 1989, A\&A, 223, 89
\reference{ECF96}Eke, V.R., Cole, S., \& Frenk, C.S.\ 1996, MNRAS, 282, 263
\reference{FaG84}Fillmore, J.A., \& Goldreich, P.\ 1984, ApJ, 281, 1
\reference{FaP94}Flores, R.A., \& Primack, J.R.\ 1994, ApJ, 427, L1
\reference{Fet88}Frenk, C.S., White, S.D.M., Davis, M., \& Efstathiou, G.P.\ 1988, ApJ, 327, 507
\reference{GaC97}Graham, A.W., \& Colless, M.M.\ 1997, MNRAS, 287, 221
\reference{GaD05}Graham, A.W., \& Driver, S.\ 2005, PASA, 22(2), 118
\reference{GETAR}Graham, A.W., Erwin, P., Trujillo, I., \& Asensio Ramos, A.\ 2003, AJ, 125, 2951
\reference{GaG03}Graham, A.W., \& Guzm\'an, R.\ 2003, AJ, 125, 2936
\reference{GLCP6}Graham, A.W., Lauer, T., Colless, M.M., \& Postman, M.\ 1996, ApJ, 465, 534
\reference{GMMDT}Graham, A.W., Merritt, D., Moore, B., Diemand, J., \& Terzi\'c, B.\ 2006a, AJ, submitted (Paper II)
\reference{GMMDT}Graham, A.W., Merritt, D., Moore, B., Diemand, J., \& Terzi\'c, B.\ 2006b, AJ, submitted (Paper II)
\reference{GTC01}Graham, A.W., Trujillo, N., \& Caon, N.\ 2001, AJ, 122, 1707
\reference{HaM06}Hansen, S.H., \& Moore, B.\ 2006, New Astronomy, 11, 333
\reference{Hen64}H\'enon, M.\ 1964, Annales d'Astrophysique, 27, 83
\reference{Her90}Hernquist, L.\ 1990, ApJ, 356, 359
\reference{Hof88}Hoffman, Y.\ 1988, ApJ, 328, 489
\reference{Jaf83}Jaffe, W.\ 1983, MNRAS, 202, 995
\reference{JaS00}Jing, Y.P., \& Suto, Y.\ 2000, ApJ, 529, L69
\reference{Kly01}Klypin, A., Kravtsov, A.V., Bullock, J.S., \& Primack, J.R.\ 2001, ApJ, 554, 903
\reference{Krav8}Kravtsov, A.V., Klypin, A.A., Bullock, J.S., \& Primack, J.R.\ 1998, ApJ, 502, 48
\reference{LNGM9}Lima Neto, G.B., Gerbal, D., \& M\'arquez, I.\ 1999, MNRAS, 309, 481
\reference{LaM01}{\L}okas, E.L., \& Mamon, G.A.\ 2001, MNRAS, 321, 155 
\reference{MaL05}Mamon, G.A., \& {\L}okas, E.L.\ 2005, MNRAS, 362, 95
\reference{Mam06}Mamon, G.A., \& {\L}okas, E.L., Dekel, A., Stoehr, F., \& Cox, T.J.\ 2006, in the 21st IAP meeting, Mass Profiles and Shapes of Cosmological Structures, ed.\ G.A. Mamon, F.\ Combes, C.\ Deffayet \& B.\ Fort (Paris: EDP) (astro-ph/0601345)
\reference{Mar00}M\'arquez, I., Lima Neto, G.B., Capelato, H., Durret, F., \& Gerbal, D.\ 2000, 353, 873
\reference{Mar01}M\'arquez, I., Lima Neto, G.B., Capelato, H., Durret, F., Lanzoni, B., \& Gerbal, D.\ 2001, A\&A, 379, 767
\reference{MaM87}Mellier, Y., \& Mathez, G.\ 1987, A\&A, 175, 1
\reference{Mer96}Merritt, D.\ 1996, AJ, 111, 2462
\reference{MA85} Merritt, D., \& Aguilar, L.~A.\ 1985, MNRAS, 217, 787 
\reference{Met05}Merritt, D., Navarro, J.F., Ludlow, A., \& Jenkins, A.\ 2005, ApJL, 624, L85
\reference{MTJ89}Merritt, D., Tremaine, S. \& Johnstone, D. 1989,
  MNRAS, 236, 829
\reference{MaT94}Merritt, D., \& Tremblay, B.\ 1994, AJ, 108, 514
\reference{Miller02} Miller, C.~J., Nichol, 
R.~C., Genovese, C., \& Wasserman, L.\ 2002, \apjl, 565, L67 
\reference{Moo94}Moore, B.\ 1994, Nature, 370, 629
\reference{Moo98}Moore, B., Governato, F., Quinn, T., Stadel, J., \& Lake, G.\ 1998, ApJ, 499, L5
\reference{Moo99}Moore, B., Quinn, T., Governato, F., Stadel, J., \& Lake, G.\ 1999, MNRAS, 310, 1147
\reference{NFW95}Navarro, J.F., Frenk, C.S., \& White, S.D.M.\ 1995, MNRAS, 275, 720
\reference{Net04}Navarro, J.F., et al.\ 2004, MNRAS, 349, 1039
\reference{NaC06}Nipoti, C, Londrillo, P., \& Ciotti, L.\ 2006, MNRAS, 370, 681
\reference{Pee70}Peebles, P.J.E.\ 1970, AJ, 75, 13
\reference{Det88}Davies, J.I., Phillipps, S., Cawson, M.G.M., Disney, M.J., \& Kibblewhite, E.J.\ 1988, MNRAS, 232, 239
\reference{PaG99}Pignatelli, E., \& Galletta, G.\ 1999, A\&A, 349, 369
\reference{PIO60}Poveda, A., Iturriaga, R., \& Orozco, I.\ 1960, Bol.\ Obs.\ Tonantzintla y Tacubaya 2, No.20, p.3
\reference{Pet05}Prada, F., Klypin, A.A., Simonneau, E., \& Betancort Rijo, J., Santiago, P., Gottl\"ober, S/ \& Sanchez-Conde, M.A.\ 2006, ApJ, 645, 1001
\reference{PaS97}Prugniel, Ph., \& Simien, F.\ 1997, A\&A, 321, 111
\reference{Ret04}Rasia, E., Tormen, G., \& Moscardini, L.\ 2004, MNRAS, 351, 237
\reference{Reed5}Reed, D., et al.\ 2005, MNRAS, 357, 82
\reference{Sco92}Scott, D.W.\ 1992, Multivariate Density Estimation, Wiley, New York
\reference{Ser63}S\'ersic, J.-L.\ 1963, Boletin de la Asociacion Argentina de Astronomia, vol.6, p.41
\reference{Ser68}S\'ersic, J.L.\ 1968, Atlas de galaxias australes
\reference{Sil86}Silverman, B.W.\ 1986, Density Estimation for Statistics and Data Analysis (Chapman and Hall: London)
\reference{Set03}Spergel, D.N.\ 2003, ApJS, 148, 175
\reference{Stadel01} Stadel, J. 2001, PhD thesis, Univ. of Washington 
\reference{SaD06}Statler, T.S., \& Diehl, S.\ 2006, BAAS, 207, 178.07
\reference{SaS95}Stepanas, P.G., \& Saha, P.\ 1995, MNRAS, 272, L13
\reference{Set01}Stiavelli, M., Miller, B.W., Ferguson, H.C., Mack, J., Whitmore, B.C., \& Lotz, J.M.\ 2001, AJ, 121, 1385
\reference{THE94}Tenjes, P., Haud, U., \& Einasto, J.\ 1994, A\&A, 286, 753
\reference{TaG05}Terzi\'c, B., \& Graham, A.W.\ 2005, MNRAS, 362, 197
\reference{TBB04}Trujillo, I., Burkert, A., \& Bell, E.F.\ 2004, ApJ, 600, 39
\reference{van61}van Albada, G.B.\ 1961, AJ, 66, 590
\reference{van82}van Albada, T.S.\ 1982, MNRAS, 201, 939
\reference{Wang05} Wang, X., Woodroofe, M., 
Walker, M.~G., Mateo, M., \& Olszewski, E.\ 2005, \apj, 626, 145 
\reference{WDO87}West, M.J., Dekel, A., Oemler, A., Jr.\ 1987, ApJ, 316, 1
\reference{Wor94} Worthey, G.\ 1994, ApJS, 95, 107
\reference{You76}Young, P.J.\ 1976, AJ, 81, 807
\reference{YaC95}Young, C.K., \& Currie, M.J.\ 1995, MNRAS, 273, 1141
\reference{Zha96}Zhao, H.S.\ 1996, MNRAS, 278, 488
\end{references}
\end{document}